\documentclass[aps,twocolumn,showpacs,preprintnumbers,nofootinbib,prd,superscriptaddress,10pt]{revtex4-2}

\makeatletter
\def\l@subsubsection#1#2{}
\def\l@subsubsubsection#1#2{}
\makeatother

\usepackage{graphicx,amssymb,amsmath,amsthm,amsfonts,epsfig,epsf}
\usepackage[usenames]{color}
\usepackage{xcolor}
\usepackage{epstopdf}
\usepackage{ulem}

\usepackage{bm}
\usepackage{amsmath}
\usepackage{dcolumn}
 \usepackage[latin1]{inputenc}
\usepackage{latexsym}
\usepackage{rotating}
\usepackage{longtable}

\usepackage{enumerate}
\usepackage{tensor,multirow}
\usepackage{mathtools}
\usepackage{url}
\usepackage[linktocpage]{hyperref}

\def\nn{\nonumber}

\newcommand{\sapienza}{Dipartimento di Fisica, Sapienza Universit\`a di Roma, Piazzale Aldo Moro 5, 00185, Roma, Italy}
\newcommand{\infn}{INFN, Sezione di Roma, Piazzale Aldo Moro 2, 00185, Roma, Italy}

\begin{document}
\title{Tidal heating as a discriminator for horizons\\ in equatorial eccentric extreme mass ratio inspirals}

\newcommand{\MPI}{Max-Planck-Institut f{\"u}r Gravitationsphysik (Albert-Einstein-Institut), D-30167 Hannover, Germany}
\newcommand{\LBNZ}{Leibniz Universit{\"a}t Hannover, D-30167 Hannover, Germany}
\newcommand{\IST}{CENTRA, Departamento de F\'{\i}sica, Instituto Superior T\'ecnico -- IST, Universidade de Lisboa -- UL, Avenida Rovisco Pais 1, 1049-001 Lisboa, Portugal}

\author{Sayak Datta}\email{sayak.datta@aei.mpg.de}
\affiliation{\MPI}\affiliation{\LBNZ}

\author{Richard Brito}
\affiliation{\IST}

\author{Scott A.\ Hughes}
\affiliation{Department of Physics and MIT Kavli Institute, MIT, Cambridge, MA 02139 USA}

\author{Talya Klinger}
\affiliation{Theoretical Astrophysics, Walter Burke Institute for Theoretical Physics, California Institute of Technology, Pasadena, California 91125, USA}

\author{Paolo Pani}

\affiliation{\sapienza}
\affiliation{\infn}

\date{\today}

\begin{abstract}
Tidal heating in a binary black hole system is driven by the absorption of energy and angular momentum by the black hole's horizon. Previous works have shown that this phenomenon becomes particularly significant during the late stages of an extreme mass ratio inspiral (EMRI) into a rapidly spinning massive black hole, a key focus for future low-frequency gravitational-wave observations by (for instance) the LISA mission.  Past analyses have largely focused on quasi-circular inspiral geometry, with some of the most detailed studies looking at equatorial cases.  Though useful for illustrating the physical principles, this limit is not very realistic astrophysically, since the population of EMRI events is expected to arise from compact objects scattered onto relativistic orbits in galactic centers through many-body events.  In this work, we extend those results by studying the importance of tidal heating in equatorial EMRIs with generic eccentricities.  Our results suggest that accurate modeling of tidal heating is crucial to prevent significant dephasing and systematic errors in EMRI parameter estimation.  We examine a phenomenological model for EMRIs around exotic compact objects by parameterizing deviations from the black hole picture in terms of the fraction of radiation absorbed compared to the BH case.  Based on a mismatch calculation we find that reflectivities as small as $|\mathcal{R}|^2 \sim \mathcal{O}(10^{-5})$ are distinguishable from the BH case, irrespective of the value of the eccentricity.  We stress, however, that this finding should be corroborated by future parameter estimation studies.
\end{abstract}
\maketitle

\section{Introduction}
\label{sec:introduction}

Black holes (BHs) within the framework of general relativity~(GR) are characterized as perfect absorbers.  This is due to their distinctive feature, the event horizon, which serves as a one-way, null hypersurface.  Detecting any degree of reflectivity in the vicinity of a dark compact object would serve as compelling evidence for deviations from the classical BH paradigm~\cite{Cardoso:2019rvt}.

Though attempting to model the reflectivity of exotic compact objects (ECOs) presents significant challenges~\cite{Oshita:2019sat,Chakraborty:2022zlq}, the absence of a horizon or the presence of adjacent structures would inevitably imply imperfect absorption.  Consequently, any examination of horizon absorption offers a model-independent means of testing the nature of compact objects and quantifying the extent to which a dark compact object can be described as a conventional BH.  

A spinning BH absorbs radiation impinging from infinity with frequency $\omega$ only when the latter surpasses a critical value,  $\omega > m\Omega_{\rm H}$, where $m$ is the azimuthal number of the incident wave and $\Omega_{\rm H} \equiv a/2Mr_+$ represents the angular velocity of the BH ($a$ is the Kerr spin parameter, and $r_+$ the coordinate radius of the event horizon).  At frequencies below this critical threshold, radiation is instead amplified by the phenomenon of superradiance~\cite{Brito:2015oca}.  In a binary system involving a pair of BHs, the radiation absorbed by the BHs has a dual description as a tidal deformation to the horizon's geometry~\cite{Hartle:1973zz, Vega2011, OSullivan:2014ywd}.  The tidal deformation then backreacts on the system, changing the binary's angular momentum and energy.  The backreaction increases the area of the holes' horizons~\cite{HawkingHartle}, increasing their entropy.  The dissipation of energy and generation of entropy leads to the apt name ``tidal heating'' for this phenomenon~\cite{Hartle:1973zz, PhysRevD.64.064004, PoissonWill}.  This effect is particularly significant in extreme-mass-ratio inspirals~(EMRIs) that are expected to be detected by space-based observatories such as the Laser Interferometer Space Antenna~(LISA)~\cite{Audley:2017drz,Colpi:2024xhw}, as it can contribute to thousands of orbital cycles if the larger supermassive object is a BH~\cite{Hughes:2001jr, Bernuzzi:2012ku, Taracchini:2013wfa, Harms:2014dqa, Datta:2019euh, Datta:2019epe,LISAConsortiumWaveformWorkingGroup:2023arg}.

\begin{figure}[t]
\includegraphics[width=\columnwidth]{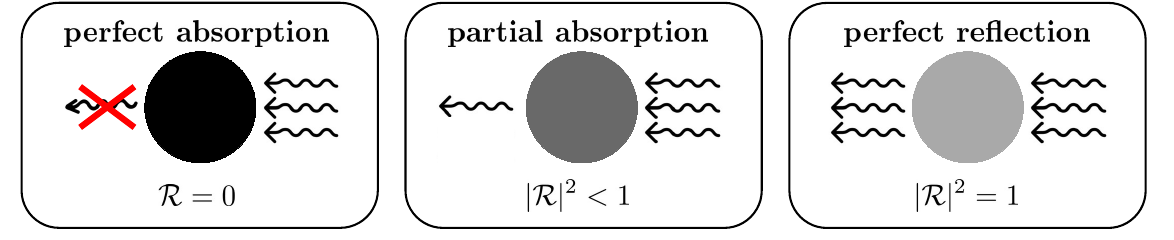}
\caption{Pictorial view of the reflectivity coefficient ${\cal R}$ to quantify the level of absorption of a compact object. The reflectivity is defined near the surface of the object, see Eq.~\eqref{eq:reflectivity}.
Left panel: Since a BH is a perfect absorber, gravitational waves impinging from the right near the horizon are absent on the left. From a one-dimensional radial perspective, there are no reflected waves, hence ${\cal R}=0$. Right panel: an object whose interaction with gravitational waves is negligible (e.g. a perfect-fluid star) has zero absorption. Perhaps counter-intuitively, from a one-dimensional radial perspective, such an object reflects all impinging radiation, so ${\cal R}=1$. Middle panel: intermediate situation in which the object is only partially absorbing.
}
\label{fig:reflectivity} 
\end{figure}

On the other hand, if one of the components in a binary system is an ECO instead of a BH, the dissipation effect is expected to be significantly reduced, possibly even to be negligible. This change can have a substantial impact on the inspiral, especially when both members of the binary are spinning rapidly, or if the larger member is spinning rapidly in an extreme mass ratio system.  Consequently, in scenarios where the external geometry of the ECO closely resembles that of a Kerr BH, tidal heating emerges as a potent model-independent tool for discerning the presence of event horizons and characterizing supermassive objects~\cite{Maselli:2017cmm, Datta:2019euh,Datta:2019epe}.

Extensive calculations are required to ascertain how tidal heating behaves within the context of particular ECO models~\cite{Pani:2010em,Macedo:2013jja,Maselli:2017cmm, Datta:2020rvo, Sherf:2021ppp}. Nonetheless, the absence of an event horizon implies a departure from the established horizon boundary conditions, which is certain to induce changes in the tidal coupling between the orbiting body and the object.  This suggests that a very useful null experiment is simply to test whether the characteristics of inspiral are consistent with the expectations of tidal heating with a BH event horizon, or whether there is a significant departure.  Previous works have shown that precise measurements with a high signal-to-noise ratio (SNR) should allow the quantification of this effect with unparalleled accuracy, whether it involves EMRIs around highly-spinning supermassive objects~\cite{Hughes:2001jr, Datta:2019euh,Datta:2019epe}, or high-spin supermassive binary systems~\cite{Maselli:2017cmm}.

Following Ref.~\cite{Datta:2019euh}, we can quantify the impact of partial absorption by establishing an upper limit on the reflectivity coefficient ${\cal R}$ of the object. The latter can be conveniently defined as being related to the fraction of energy that is lost inside the object compared to the BH case. Schematically,
\begin{equation}
    \left(\frac{dE}{dt}\right)^{\text{tot}} = \left(\frac{dE}{dt}\right)^{\infty} + (1 - |\mathcal{R}|^2)\left(\frac{dE}{dt}\right)^{H},
    \label{eq:reflectivity}
\end{equation}
%%%
where $\left(\frac{dE}{dt}\right)^{\infty}$ is the energy flux at infinity, whereas $(1 - |\mathcal{R}|^2)\left(\frac{dE}{dt}\right)^{H}$ is the amount of energy dissipated within the object per unit time. For a BH, ${\cal R}=0$ and this quantity reduces to the energy flux across the horizon, $\left(\frac{dE}{dt}\right)^{H}$, associated to tidal heating. For a perfectly reflecting object, $|{\cal R}|=1$ and this contribution is absent.
%%%
Note that the term reflectivity is associated to the one-dimensional radial description of the scattering, wherein zero reflectivity means total absorption while perfect reflectivity means that waves do not interact with the object, see Fig.~\ref{fig:reflectivity} for a pictorial view.\footnote{Our reflectivity coefficient should not be confused with the reflectivity defined at infinity in a scattering process~\cite{Starobinskij2}. In that case the BH reflectivity is associated to the scattering of waves off the effective potential, namely the BH greybody factor. The reflectivity defined in this work is the same concept but applied near the compact object, so that it accounts only for the possible interaction of radiation with the interior of the object.}

To our knowledge, the impact of what we call tidal heating was first carefully examined in Ref.\ \cite{Poisson:1994yf}.  Focusing on circular and equatorial orbits of Schwarzschild black holes, Poisson and Sasaki showed that the down-horizon flux is a tiny fraction of the flux to infinity, scaling with orbital speed $\propto v^8$ and with a very small coefficient.  They concluded that this term would make a negligible contribute to wave phasing for (then planned) gravitational-wave observations.  Later work \cite{Tagoshi:1997jy} confirmed a result from Gal'tsov \cite{Galtsov:1982hwm} that the horizon term is significantly stronger for orbits of Kerr black holes, changing the scaling from $v^8$ to a form related to the horizon's rotation frequency times $v^5$.  Motivated by this form, Ref.\ \cite{Hughes:2001jr} examined the impact of these fluxes on quasi-circular inspiral, finding that the effect of tidal heating could change by many thousands the number of orbits a secondary executes in the LISA band, especially for inspiral into a rapidly spinning black hole.  This result was examined very carefully for equatorial, quasi-circular EMRIs in~\cite{Datta:2019epe} (see also Refs.~\cite{Isoyama:2017tbp, Datta:2020gem, Mukherjee:2022wws, Lyu:2023zxv} for the case of stellar-mass binaries).  Later the work was extended including the impact of resonances~\cite{Sago:2021iku, Maggio:2021uge, Cardoso:2022fbq} and of modified boundary conditions arising from the object interior.

In this paper, our primary objective is to extend the results of Ref.\ \cite{Datta:2019epe} in order to study equatorial eccentric orbits.  Lifting the restriction to quasi-circularity is particularly important: astrophysical EMRIs are expected to be dominated by binaries created through multibody scattering events, so substantial eccentricity is likely to be the norm for these events.  Understanding how eccentricity complicates the picture is important for assessing how well the null experiment which tests of tidal heating make possible can be performed.  Our specific aim is to examine the importance of tidal heating for eccentricities that cover a range consistent with what is expected for astrophysical systems, and to assess at least roughly how well the reflectivity of a spinning supermassive object can be constrained by measuring tidal heating with EMRI gravitational waves (GWs).

We call particular attention to a parallel analysis by Zi, Ye, and Li~\cite{Zi:2023geb} which appeared as we were completing this analysis.  Their work covers much the same ground as the scenarios that we study.  Indeed, they go beyond our analysis by examining orbit configurations that are both eccentric and inclined.  The most important point of difference is that their analysis uses analytic flux formulas and waveforms based on an update of the ``augmented analytic kludge'' (AAK) \cite{Chua:2017ujo} implemented in the Fast EMRI Waveform model \cite{Katz2021}.  This model produces waveforms much more rapidly than the framework we have developed, which it makes it possible for the authors of \cite{Zi:2023geb} to develop Fisher-information-matrix based estimates of parameter measurement accuracy.  However, it must be noted that the AAK model is least accurate in the strong-field, fast-motion domain in which the effects of tidal heating are most important.  Their work provides a very valuable quantitative first study of how well the physics of tidal heating may be probed with future observations; we expect the fluxes and waveforms we have developed to be reliable deep in the strong field, but cannot yet be used for detailed parameter studies of the sort studied in \cite{Zi:2023geb}.  As the Fast EMRI Waveform framework is expanded to cover the Kerr parameter space using black-hole-perturbation-theory-based waveforms, it should not be difficult to revisit the analysis of \cite{Zi:2023geb} using strong-field waveforms.

Throughout this article, we use $G = 1 = c$ units. The mass and angular momentum of the primary object are denoted by $M$ and $J = aM = \chi M^2$, respectively. The mass of the small orbiting (nonspinning) body is denoted by $\mu$, and the mass ratio is denoted by $\eta = \mu/M \ll 1$.

\section{Setup}
In this section, we summarize how to compute an adiabatic inspiral and its associated gravitational waveform.  Since comprehensive details are available in the references (see Ref.\ \cite{Hughes:2021exa} in particular), this brief overview aims to concisely introduce critical quantities and concepts necessary for this analysis.
\subsection{Background metric and orbits}
\label{sec:backgroundandorbits}

We take the primary object to be a Kerr BH~\cite{Kerr1963}, in keeping with the idea that our goal is to formulate a test of the hypothesis that these objects have an event horizon as the Kerr metric requires~\cite{Maggio:2017ivp, Abedi:2016hgu, Wang:2018gin, Barausse:2018vdb, Datta:2019epe}.  If measurements were to find some deviation from the Kerr hypothesis, it would be an interesting question what drives this deviation.  For instance, there could be some systematic error in the measurement model, or a neglected environmental effect.  Perhaps most interesting is the idea that the ``massive compact object'' is something other than GR BH.  Given the number of ways that one could imagine deviating from this ``standard'' hypothesis, our view is that the most natural starting point is to formulate a way of assessing this hypothesis, and let data and Nature determine whether any modifications are necessary.

Consequently, our analysis starts by assuming that the small object moves on an orbit of the Kerr metric.  We work in the adiabatic limit~\cite{Hughes:2021exa}, imagining that these orbits secularly evolve due to the backreaction of GW emission.  We begin with the Kerr metric in Boyer-Lindquist coordinates~\cite{BoyerLindquist1967}:

\begin{align}
ds^2&=-\left(1-\frac{2Mr}{\Sigma}\right)dt^2+\frac{\Sigma}{\Delta}dr^2-\frac{
4Mr}{\Sigma}a\sin^2\theta d\phi dt  \nn \\
&+{\Sigma}d\theta^2+
\left[(r^2+a^2)\sin^2\theta +\frac{2Mr}{\Sigma}a^2\sin^4\theta
\right]d\phi^2\,,\label{Kerr}
\end{align}
where $\Sigma = r^2 + a^2\cos^2\theta$ and $\Delta = r^2 - 2Mr + a^2 = (r - r_+)(r - r_-)$, with $r_\pm = M \pm \sqrt{M^2 - a^2}$. The angular velocity at the event horizon is $\Omega_H = a/(2Mr_+)$.  Note that the Boyer-Lindquist coordinate $t$ is the time parameter used by observers very far from the BH.  As such, the evolution of observables is very naturally parameterized using $t$.

Our analysis is based on an adiabatic approximation to inspiral, in which the smaller body's motion is treated as a geodesic orbit which slowly evolves due to the backreaction of GW emission.  The detailed properties of Kerr geodesic orbits are described in Refs.~\cite{schmidt2002, Fujita:2009bp}.  We focus on orbits confined to the equatorial plane, meaning that their inclinations are either $I = 0^\circ$ (prograde) or $I = 180^\circ$ (retrograde).  Equatorial orbits are parameterized up to initial conditions by the integrals of motion $E$ (energy per unit mass) and $L_z$ (axial angular momentum per unit mass), or by the orbital geometry parameters $p$ (semi-latus rectum) and $e$ (eccentricity).  (A third integral of the motion, the Carter constant $Q$, is zero for equatorial orbits, and so plays no role in our present analysis.)  These parameters specify the orbit's radial motion:
\begin{equation}
    r = \frac{p}{1 + e\cos\chi_r}\;.
\end{equation}
As the angle $\chi_r$ varies from $0$ to $\pi$ to $2\pi$, the position of the smaller body oscillates from periapsis $p/(1 + e)$ to apoapsis $p/(1 - e)$ and back.  It is straightforward to recast the radial geodesic equation as an evolution equation for $\chi_r$; see \cite{Fujita:2009bp}.  The parameterizations $(E, L_z)$ and $(p, e)$ are entirely equivalent; formulas exist allowing one to switch between these forms with ease \cite{schmidt2002, Fujita:2009bp, vandemeent2020, Hughes2024}.

\subsection{Linear perturbations from the secondary}
\label{sec:perturbations}

We describe the impact of the secondary on the binary using the Teukolsky equation~\cite{Teukolsky:1973ha}, which describes perturbations to the curvature of Kerr BHs.  We focus on the Newman-Penrose curvature scalar $\psi_4$, which describes radiative components of the Weyl curvature tensor~\cite{Newman:1961qr}.  The equation governing $\psi_4$ has the form
\begin{equation}
    \mathcal{D}^2\psi_4 = 4\pi\Sigma\mathcal{T}\;,
    \label{eq:teuk_schematic}
\end{equation}
where $\mathcal{D}^2$ is a second-order linear differential operator whose precise form is not important for our purposes, and $\mathcal{T}$ is a source that is constructed from the stress-energy tensor describing a point-like body moving in the Kerr spacetime.  See, for example, Refs.\ \cite{Teukolsky:1973ha, Hughes:2021exa} for detailed discussion of $\mathcal{D}^2$ and $\mathcal{T}$.

Far from the source, the scalar $\psi_4$ is given by the system's GWs:
\begin{equation}
\label{eq: waveform psi4 relation}
    \psi_4 = \frac{1}{2}\frac{d^2}{dt^2}(h_+ - ih_\times) \quad \text{as} \quad r \rightarrow \infty.
\end{equation}
From this limiting form of $\psi_4$, one can extract the rate at which GWs carry energy $E$ and axial angular momentum $L_z$ from the source.  As $r \to r_+$, $\psi_4$ also encodes information about radiation absorbed by the BH, or equivalently, tidal interactions between the orbiting body and the event horizon.  From this, one can likewise extract the rate at which energy and axial angular momentum are exchanged between the secondary's orbit and the primary BH.  The behavior of $\psi_4$ as $r \to r_+$ is thus crucial for understanding tidal heating in BH binary systems.  Knowledge of $\psi_4$ in the limits $r\to \infty$ and $r\to r_+$ provides the data we use for constructing adiabatic inspirals.

To solve the Teukolsky equation, we first separate $\psi_4$ by introducing the Fourier and multipolar expansion
\begin{widetext}
    \begin{equation}
    \psi_4 = \frac{1}{(r - ia\cos\theta)^4}\int_{-\infty}^{\infty}d\omega\sum_{l=2}^{\infty}\sum_{m=-l}^{l}R_{lm}(r;\omega)S_{lm}(\theta;a\omega)e^{i[m\phi - \omega (t - t_0)]}.
\end{equation}
\end{widetext}
This expansion separates Eq.~\eqref{eq:teuk_schematic}, with ordinary differential equations governing $R_{lm}(r, \omega)$ and $S_{lm}(\theta, a\omega)$~\cite{Teukolsky:1973ha}.  Note that, following Ref.\ \cite{Hughes:2021exa}, we have introduced an initial time $t_0$ in anticipation of this formula's later application to inspirals.

The separated radial function has the following asymptotic behavior:
\begin{equation}
    R_{lm}(r, \omega) \rightarrow \begin{cases}
        Z^{\infty }_{lm\omega}r^3e^{i\omega r^*}, & \text{as } r \rightarrow \infty, \\
        Z^{H }_{lm\omega}\Delta e^{-i(\omega - m\Omega_H)r^*}, & \text{as } r \rightarrow r+,
    \end{cases}
\end{equation}
where $r^*$ is the tortoise coordinate defined through the relation $dr_*/dr=(r^2+a^2)/\Delta$.  For equatorial orbits, the coefficients $Z^{\infty,H}_{lm\omega}$ can be further decomposed as
\begin{equation}
    Z^{\infty,H}_{lm\omega} = \sum_{n=-\infty}^{\infty}Z^{\infty,H}_{lmn}\delta(\omega - \omega_{mn}),
\end{equation}
where we have introduced
\begin{equation}
    \omega_{mn} = m\Omega_\phi + n\Omega_r\;.
\end{equation}
The frequency $\Omega_\phi = 2\pi/T_\phi$, where $T_\phi$ is the Boyer-Lindquist time interval for the orbiting body to move through $2\pi$ radians of axial angle $\phi$.  The frequency $\Omega_r = 2\pi/T_r$, where $T_r$ is the Boyer-Lindquist time interval for the orbiting body to complete a full cycle of orbital motion (e.g., for the body to move in its eccentric orbit from apoapsis to peripasis and back).  The coefficients $Z^{\infty,H}_{lmn}$ are found by integrating a Green's function, constructed from homogeneous solutions to the separated radial equation, against the Teukolsky equation's source term.  For further details, see Ref.\ \cite{Hughes:2021exa}.

For our purposes, the key thing to emphasize is that once $Z^{\infty,H}_{lmn}$ are known, the function $\psi_4$ is completely known in the limits $r\to\infty$ and $r\to r_+$.  For example,
\begin{equation}
    \psi_4(r\to\infty) = \frac{1}{r}\sum_{lmn} Z^\infty_{lmn}S_{lm}(\theta;a\omega_{mn})e^{i[m\phi - \omega_{mn}(t - t_0)]}\;.
    \label{eq:psi4_final}
\end{equation}
Formally, the sum over $l$ goes from $2$ to $\infty$; the $m$ sum from $-l$ to $l$; and the $n$ sum from $-\infty$ to $\infty$.  In practice, these sums are truncated after certain convergence criteria, discussed in Sec.\ \ref{sec:inspiral}, are met.  The corresponding form for $\psi_4(r \to r_+)$ is significantly more complicated; we refer the reader to Ref.\ \cite{OSullivan:2014ywd} for detailed discussion of how this solution is constructed and used to compute tidal coupling and the down horizon fluxes.  Once the set of coefficients $Z^{\infty,H}_{lmn}$ has been computed, we have all the information we need to compute adiabatic backreaction and its associated gravitational waveform.

\begin{figure*}
\includegraphics[width=89mm]{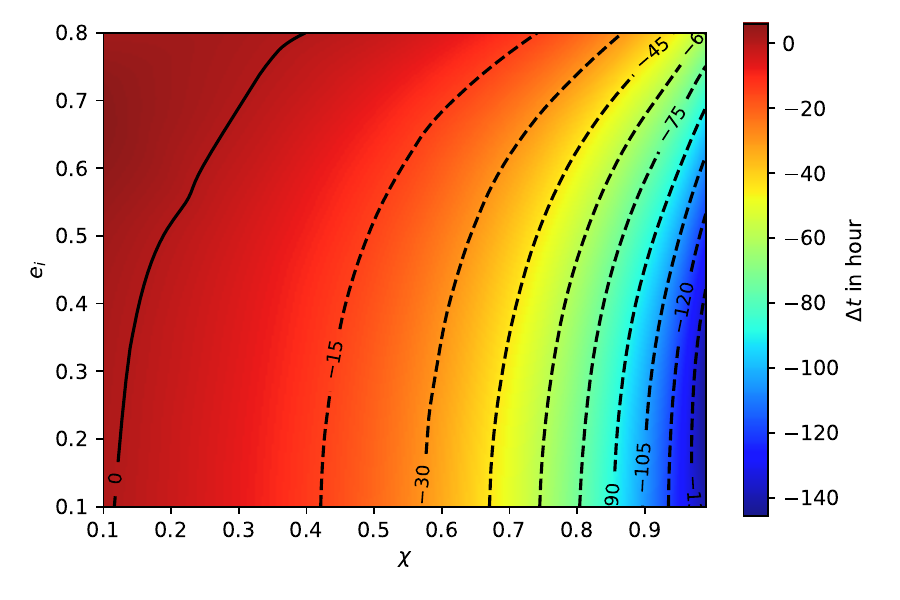}
\includegraphics[width=89mm]{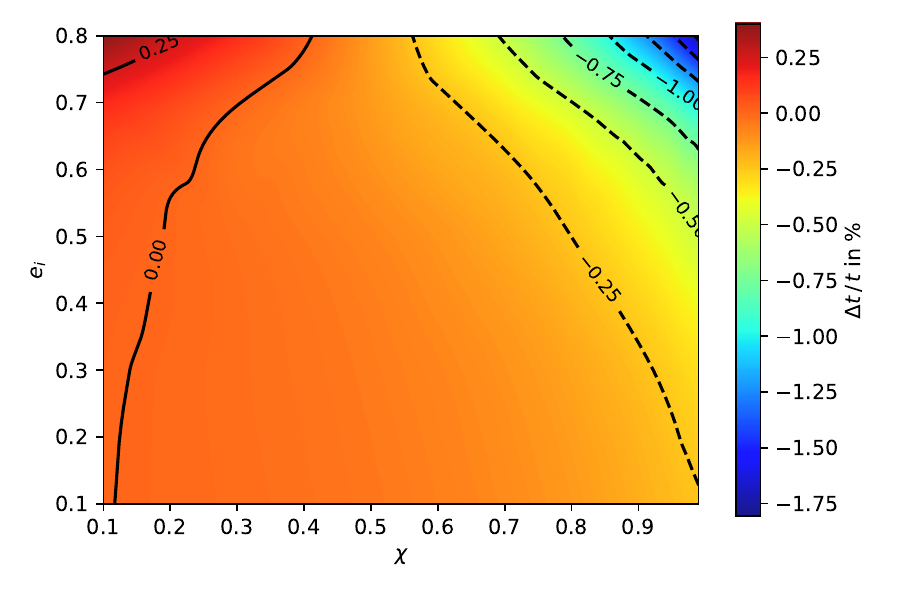}
\includegraphics[width=89mm]{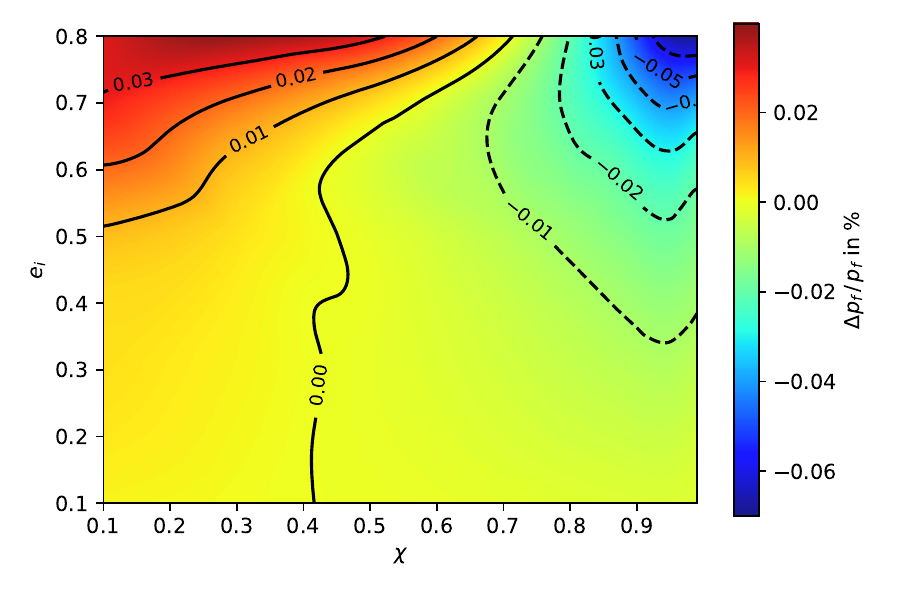}
\includegraphics[width=89mm]{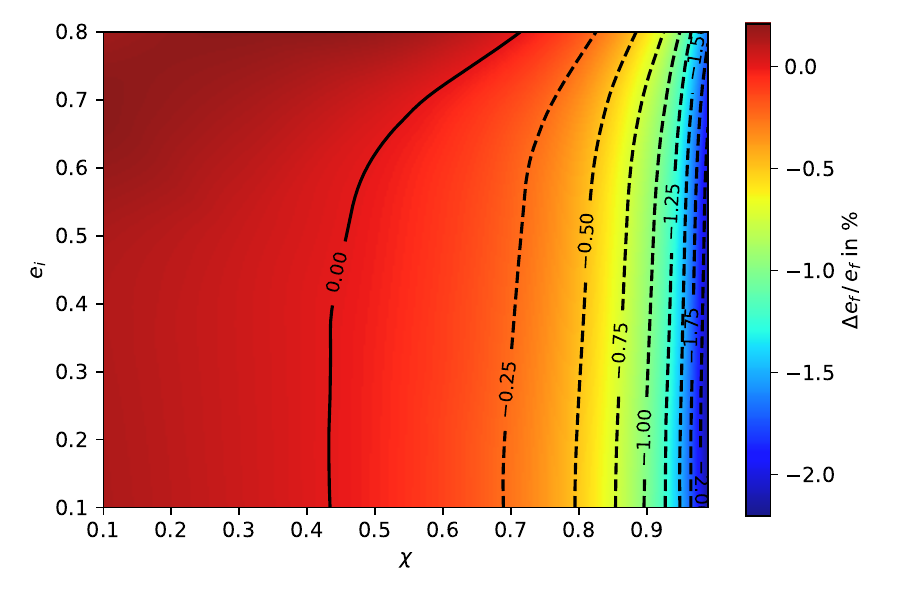}
\caption{Maximum change to the inspiral trajectory arising due to tidal heating.  In these four panels, we examine a variety of prograde inspirals; in all cases, the larger BH has mass $10^6\,M_\odot$, and the system has mass ratio $M/\mu = 3\times10^4$.  We vary the spin $\chi$ of the larger BH (horizontal axis of the figures), as well as the initial eccentricity $e_i$.  We choose the initial semi-latus rectum $p_i$ such that the orbital harmonic $\omega_{20} = 2\pi \times 10^{-3}\,{\rm sec}^{-1}$.  (The $m = 2$, $n = 0$ harmonic typically makes the loudest contribution to the gravitational waveform.)  The four panels compare inspirals computed in ``normal'' GR ($\mathcal{R} = 0$) with maximum reflectivity ($\mathcal{R} = 1$); the various changes we compute scale with the value of $|\mathcal{R}|^2$.  Top left shows how the duration (which varies from $0.8$ months to 81.9 months across this parameter space; our ``month'' is 30 days) changes with reflectivity; top right shows similar information, but presented as a fraction of the total inspiral duration.  Both panels indicate that, across much of the parameter space, tidal heating makes a non-negligible change to the inspiral duration, just as previous work indicated for quasi-circular inspiral.  The bottom panels quantify the extent to which the path in the $(p,e)$ plane is affected by tidal heating, with bottom left showing how the value of $p$ at which inspiral ends changes, and bottom right showing similar data for the final value of $e$.  The change to the trajectory is a qualitatively new aspect to the problem seen in the eccentric case.  By contrast, for quasi-circular inspiral, all systems evolve with $e = 0$ through a fixed range of orbital radius.  This illustrates how the physics of these systems are complicated when we consider more realistic orbital configurations.}
\label{fig:orbit change pro} 
\end{figure*}

\subsection{Gravitational waveform for an adiabatic inspiral}
\label{sec:inspiral}

Combining Eqs.\ (\ref{eq: waveform psi4 relation}) and (\ref{eq:psi4_final}) yields the following expression for the gravitational waveform from an extreme mass ratio system:

\begin{align}
\label{eq: h A connection}
    h_+ - ih_\times &\equiv \frac{1}{r}\sum_{lmn}h_{lmn}
    \nonumber\\
    &= \frac{1}{r}\sum_{lmn}A_{lmn}S_{lm}(\theta; a\omega_{mn})e^{i[m\phi - \omega_{mn}(t - t_0)]}\;,
\end{align}
where

\begin{equation}
    A_{lmn} = -2Z^{\infty}_{lmn}/\omega_{mn}^2\;.
\end{equation}
To describe the gravitational waveform associated with an inspiral, we need to promote various terms in Eq.\ (\ref{eq: h A connection}) to quantities which evolve during inspiral.  Before doing so, we first describe how we compute this inspiral.

\begin{figure*}
\includegraphics[width=89mm]{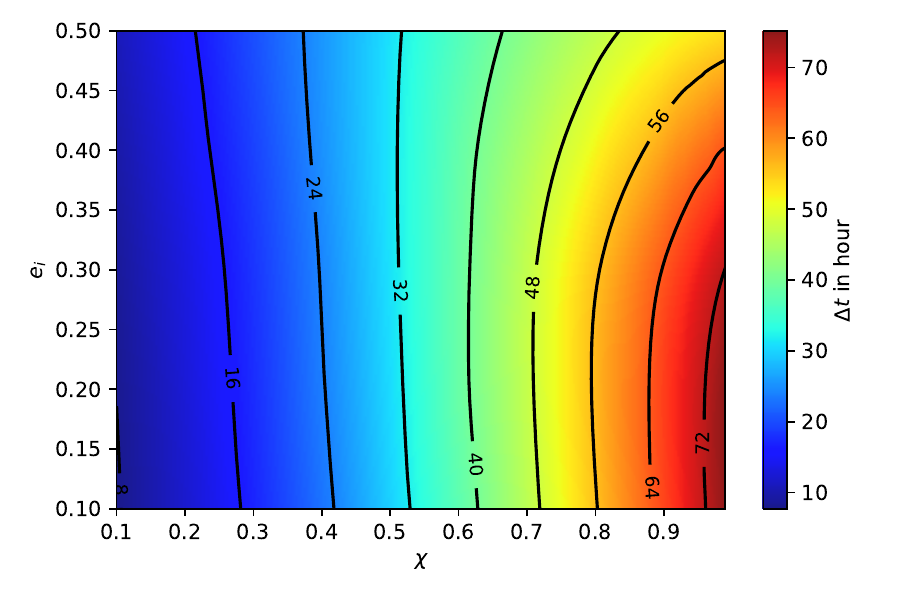}
\includegraphics[width=89mm]{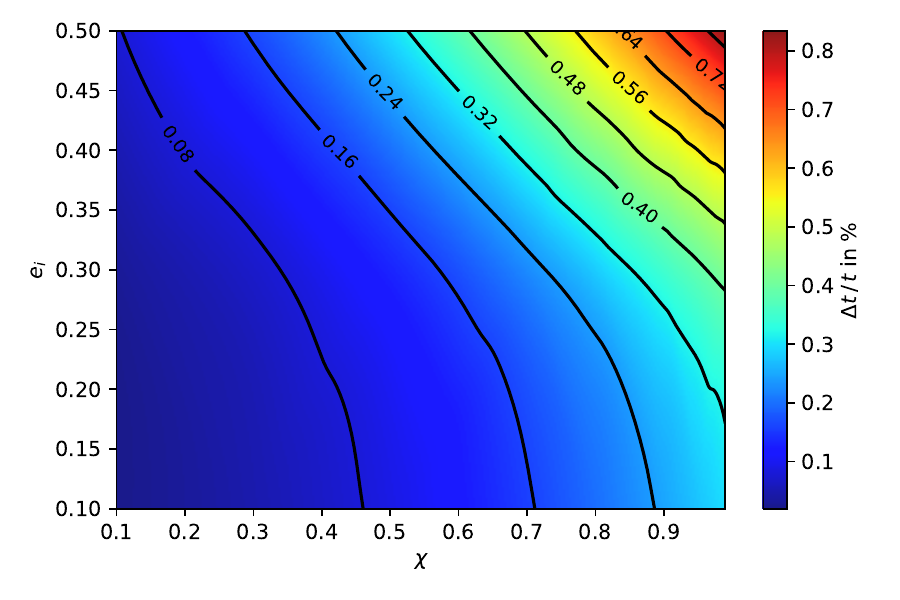}
\includegraphics[width=89mm]{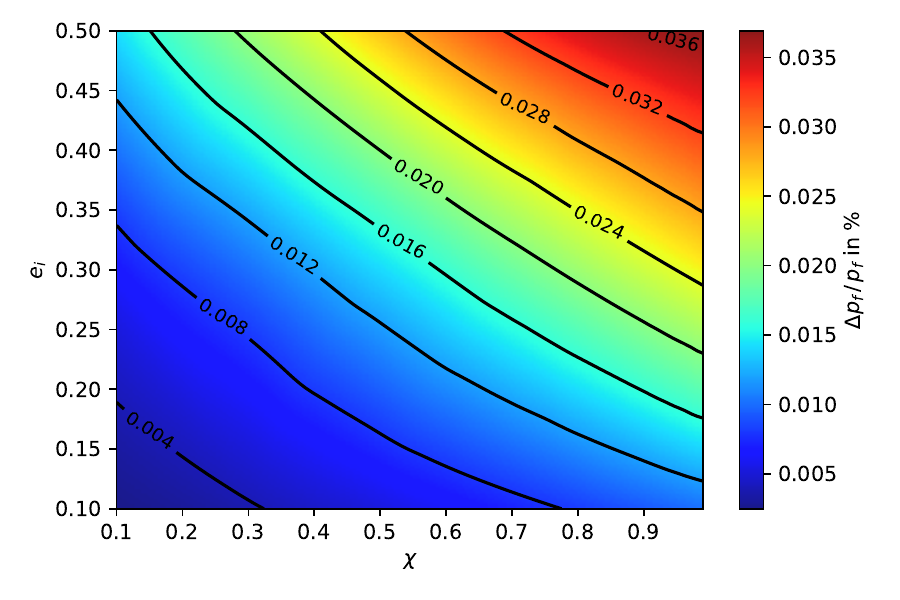}
\includegraphics[width=89mm]{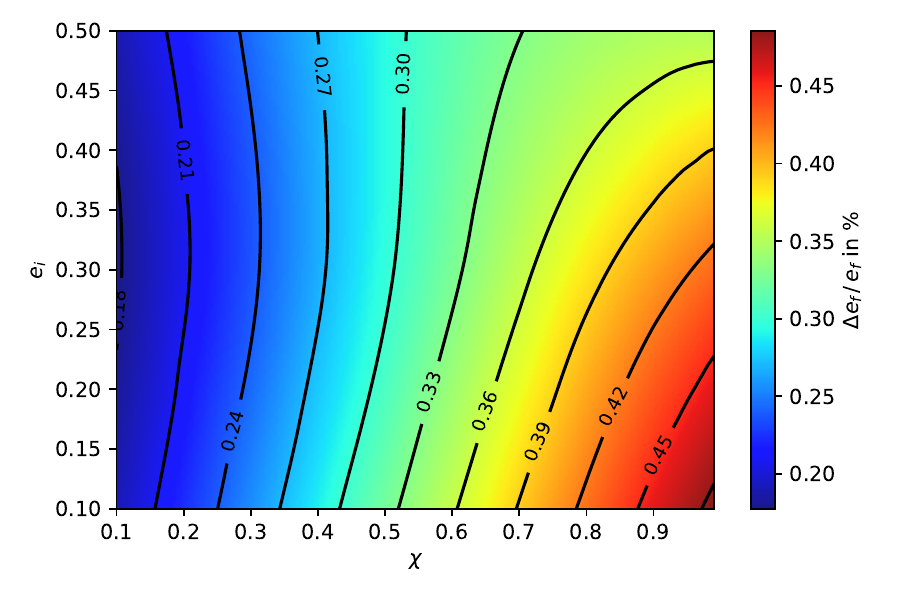}
\caption{Maximum change to the inspiral trajectory arising due to tidal heating.  This figure shows the same information as Fig.\ \ref{fig:orbit change pro}, but now considers retrograde orbits.  The total duration of inspirals in this case varies from 8.8 to 55.3 months across this parameter space.  Although similar trends are seen as those found in the prograde examples, the changes due to tidal heating are typically much smaller here.  This is not surprising: the last stable orbit is typically at much wider separation for retrograde orbits, so the influence of tidal heating (which is strongest as orbits come close to the event horizon) tends to be smaller for these orbits.}
\label{fig: orbit change ret} 
\end{figure*}

\subsubsection{Adiabatic Backreaction}

We model adiabatic inspiral by allowing orbits to evolve in response to the backreaction of GWs.  At the adiabatic level, backreaction is equivalent to computing the action of the orbit-averaged dissipative part of the gravitational self force.  Its impact can be expressed using the amplitudes $Z^{\infty,H}_{lmn}$ that we find by solving the separated Teukolsky equation to describe the rate of change of the orbit integrals (see, e.g., Ref.\ \cite{Hughes:2021exa} for summary discussion):
\begin{align}
    \left(\frac{dE}{dt}\right)^\infty &= \sum_{lmn}\frac{|Z^{\infty}_{ lmn}|^2}{4\pi\omega_{mn}^2}\;,
    \label{eq:dEIdt}\\
    \left(\frac{dE}{dt}\right)^H &= \sum_{lmn}\alpha_{lmn}\frac{|Z^{H}_{ lmn}|^2}{4\pi\omega_{mn}^2}\;,
    \label{eq:dEHdt}\\
    \left(\frac{dL_z}{dt}\right)^\infty &= \sum_{lmn}\frac{m|Z^{\infty}_{ lmn}|^2}{4\pi\omega_{mn}^3}\;,
    \label{eq:dLzIdt}\\
    \left(\frac{dL_z}{dt}\right)^H &= \sum_{lmn}\alpha_{lmn}\frac{m|Z^{H}_{ lmn}|^2}{4\pi\omega_{mn}^3}\;,
    \label{eq:dLzHdt}
\end{align}
where the superscripted $\infty$ and $H$ represent contributions to the evolution of these orbit integrals from radiation to infinity and from the interaction with the larger BH's event horizon, respectively.  The detailed form of the coefficient $\alpha_{lmn}$ can be read out of Eqs.\ (3.30)--(3.32) of Ref.~\cite{Hughes:2021exa}.  Adiabatic inspiral is then enacted by enforcing global conservation: letting $\mathcal{C}$ stand for either $E$ or $L_z$,

\begin{equation}
    \left(\frac{d\mathcal{C}}{dt}\right)^{\text{orbit}} = -\left(\frac{d\mathcal{C}}{dt}\right)^{\infty} - \left(\frac{d\mathcal{C}}{dt}\right)^{H}\;.
\end{equation}

These balance laws assume that the EMRI system adapts to changes in the orbital integrals while maintaining a geodesic trajectory by adjusting the orbit's geometry accordingly.  This allows us to relate the fluxes to the rate-of-change of the orbit parameters $p$ and $e$ discussed in Sec.\ \ref{sec:backgroundandorbits}. We follow the method of Ref.~\cite{Hughes:2021exa} to extract the changes in orbital parameters from the fluxes.

The fluxes at the horizon directly encode the ``tidal heating'' that is the focus of this study.  In order to study the importance of these terms and to formulate the test we propose, we consider a phenomenological model\footnote{Note that an imperfect absorption would give rise to modified boundary conditions affecting one of the solutions to the homogeneous Teukolsky equation which, in turn, would affect both losses at infinity and at the horizon which would modify the balance law \cite{Maggio:2021uge}. In the spirit of a model independent test, we opted here for the minimal change in Eq.\ \eqref{eq:modifiedbalance}, where ${\cal R}$ is a constant.  More accurate models would provide explicit frequency-dependent expressions ${\cal R}(\omega)$, and would modify Eq.\ \eqref{eq:modifiedbalance} consistently.} in which we modify this balance equation to

\begin{equation}
    \left(\frac{d\mathcal{C}}{dt}\right)^{\text{orbit}} = -\left(\frac{d\mathcal{C}}{dt}\right)^{\infty} - (1 - |\mathcal{R}|^2)\left(\frac{d\mathcal{C}}{dt}\right)^{H},
    \label{eq:modifiedbalance}
\end{equation}
where $\mathcal{R}$ is a reflectivity coefficient (see Fig.~\ref{fig:reflectivity} and Eq.~\eqref{eq:reflectivity}).  When $\mathcal{R} = 0$, we recover the standard GR adiabatic inspiral; otherwise, the horizon flux is modified, changing the inspiral.  Our goal in the analysis that follows is to study and understand this modification.

\subsubsection{Adiabatic inspiral along a sequence of geodesics}

Turn now to how we build the waveform for a source that is slowly inspiraling according to the balance law (\ref{eq:modifiedbalance}).  Using this law, it is straightforward to integrate from some initial time $t_0$ until the moment the small body reaches the last stable orbit (LSO) and stable geodesics no longer exist, thereby constructing the orbit integrals $E(t_i)$ and $L_z(t_i)$ as functions of a ``bookkeeper'' inspiral time $t_i$.  Combining the functions with the constraint that the orbit must evolve from geodesic to geodesic allows us to reparameterize~\cite{Hughes2024} and obtain the parameters that describe the orbit's geometry at every moment, the semi-latus rectum $p$ and orbital eccentricity $e$.  Finally, with $p(t_i)$ and $e(t_i)$ in hand, we can compute the orbit's frequecies $\Omega_{r,\phi}(t_i)$ and the Teukolsky function amplitude $Z^\infty_{lmn}(t_i)$ at each moment of inspiral.  We then ``upgrade'' the curvature scalar (\ref{eq:psi4_final}) to the form

\begin{figure*}
\includegraphics[width=89mm]{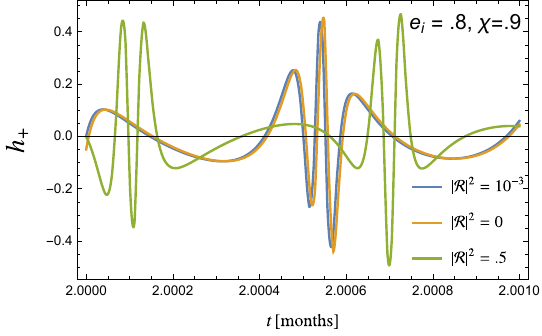}
\includegraphics[width=89mm]{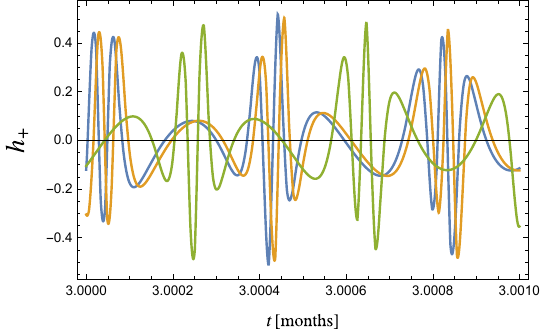}
\caption{Two moments along an example waveform, computed for three different reflectivities.  Both panels show waveforms corresponding to a prograde inspiral into a BH with $\chi = 0.9$, for a system with initial eccentricity $e = 0.8$; other parameters are as described in the caption to Fig.\ \ref{fig:orbit change pro}.  The left panel shows a span of $0.001$ month, or roughly 45 minutes starting 2 months into inspiral; the right panel shows the same thing, but 3 months into inspiral.  Notice that at two months, the case with $|\mathcal{R}|^2 = 0.5$ is highly dephased from the other two examples; after three months, even the case with $|\mathcal{R}|^2 = 10^{-3}$ is noticeably dephased from the GR waveform ($\mathcal{R} = 0$).}

\label{fig: waveform} 
\end{figure*}

\begin{equation}
    \psi_4(t_i) = \frac{1}{r} \sum_{lmn} Z^\infty_{lmn}(t_i) S_{lm}[\theta; a\omega_{mn}(t_i)] e^{i[m\phi - \Phi_{mn}(t_i)]}.
    \label{eq:psi4_inspiral}
\end{equation}
This expression is similar to Eq.\ (\ref{eq:psi4_final}), but the amplitude $Z^\infty_{lmn}$ and frequency $\omega_{mn}$ are now functions of $t_i$, and the term $\omega_{mn}(t - t_0)$ in the exponential has been replaced by the accumulated phase
\begin{equation}
    \Phi_{mn}(t_i) = \int_{t_0}^{t_i} \omega_{mn}(t') \, dt'\;.
\end{equation}
This phase reduces to $\omega_{mn}(t_i - t_0)$ for a non-evolving system, reproducing Eq.\ (\ref{eq:psi4_final}) in that limit.  Substituting (\ref{eq:psi4_inspiral}) into the left-hand side of the Teukolsky equation (\ref{eq:teuk_schematic}), one finds that it solves this equation up to errors $\mathcal{O}(\eta)$.  These errors are in turn due to the fact that time derivatives have ``fast-time'' contributions from terms that vary on orbital timescales (i.e., on timescales $\sim T_r$ or $\sim T_\phi$, both of which are $\sim M$), as well as ``slow-time'' contributions from terms that vary on the inspiral timescale ($\sim M/\eta$).  The adiabatic approximation neglects the slow-time contribution to various time derivatives, incurring this error.  Post-adiabatic waveforms which are currently under development will ultimately be needed to correct these leading waveform errors \cite{Wardell2023}.

Our primary interest is in the waveform $h(t_i)$, which is given by
\begin{equation}
    h(t_i) \equiv \sum_{lmn} h_{lmn}(t_i) = \frac{1}{r} \sum_{lmn} H_{lmn}(t_i) e^{i[m\phi - \Phi_{mn}(t_i)]},
\end{equation}
where
\begin{equation}
    H_{lmn}(t_i) = A_{lmn}(t_i) S_{lm}[\theta; a\omega_{mn}(t_i)]
\end{equation}
and
\begin{equation}
    A_{lmn}(t_i) = -\frac{2 Z^\infty_{lmn}(t_i)}{\omega_{mn}(t_i)^2}.
\end{equation}
We use numerical data describing waveforms computed in this manner to study the impact of modifications to tidal heating on these systems. 

\section{Numerical implementation}

Here we briefly summarize important details about our numerical implementation of adiabatic inspiral and EMRI waveform computation.  We begin by computing $Z^{\infty,H}_{lmn}$ on a grid that covers a large section of the $(p, e)$ plane.  The grid we use is closely related to the one that was developed for the ``Fast EMRI Waveform'' (FEW) project \cite{Chua:2020stf, Katz2021}.  For each spin value that we examine, our grid has 40 points in $e$ over the domain $0 \le e \le 0.8$, evenly spaced in $e^2$ to yield denser coverage at small eccentricity.  We use 36 points in the $p$ direction, spaced according to the formula
\begin{equation}
    p_j = p_{\rm min} + 4M(e^{j\Delta u} - 1)\;,\quad 0\le j \le 35\;.
\end{equation}
We use $\Delta u = 0.035$, and set $p_{\rm min} = p_{\rm LSO} + 0.05M$; the LSO can be computed very accurately as a function of BH spin $a$ and eccentricity $e$ using Ref.\ \cite{Stein:2019buj}.  We offset the inner edge of our grid slightly from the LSO because certain fields vary quite rapidly over the domain $p_{\rm LSO} \lesssim p \lesssim p_{\rm LSO} + 0.05M$.  At the mass ratios relevant to EMRI systems, backreaction becomes so strong that the adiabatic description of inspiral begins to break down somewhere near the inner edge of our grid.  Truncating our grids with an offset of $0.05M$ from the LSO allows us to focus our analysis on the domain where the adiabatic description is most likely to be highly reliable.  Very little inspiral remains by the time the small body reaches $p_{\rm min}$, so any errors incurred by truncating at $p_{\rm min}$ rather than closer to $p_{\rm LSO}$ are negligible.  We take the large member of the EMRI to have mass $M = 10^6\,M_\odot$, and fix the secondary's mass such that $M/\mu = 3 \times 10^4$.  Many of the critical quantities we compute in this analysis scale with these masses in a simple way, so it is not difficult to rescale to other configurations.

\begin{figure*}
\includegraphics[width=89mm]{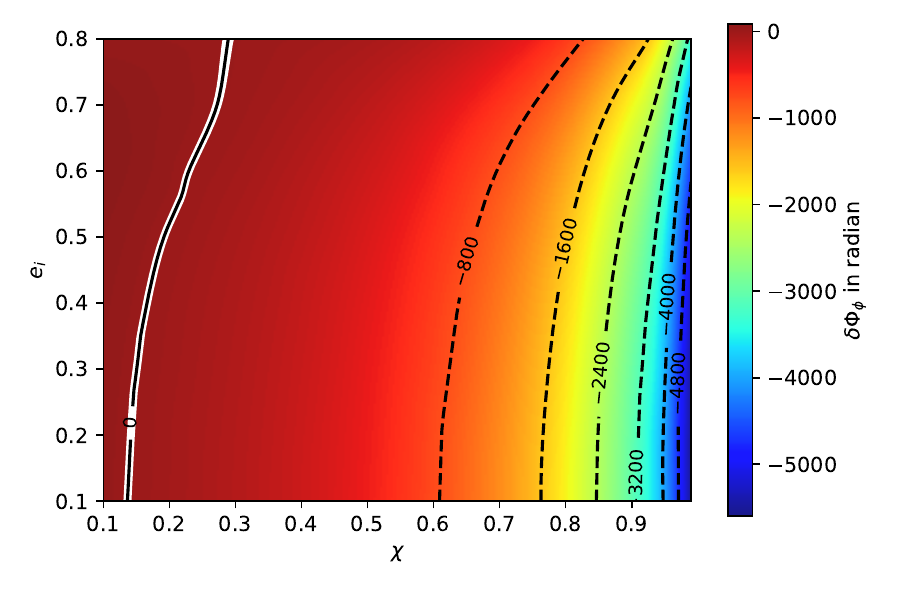}
\includegraphics[width=89mm]{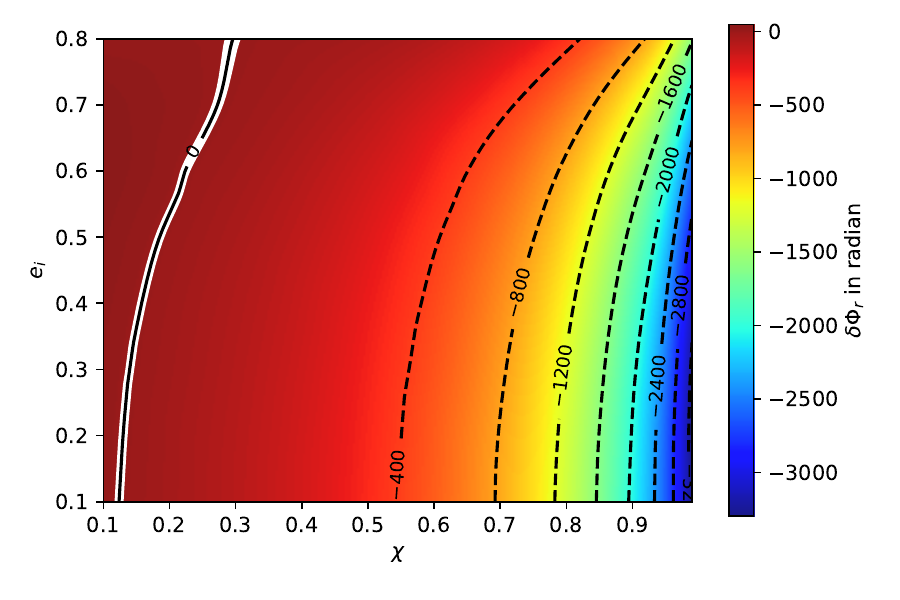}

\caption{Total dephasing found as a function of $\chi$ and $e_i$ for prograde inspirals, comparing the ``normal'' GR inspiral ($\mathcal{R} = 0$) to maximum reflectivity ($\mathcal{R} = 1$).  The binaries considered are the same ones described in the caption to Fig.\ \ref{fig:orbit change pro}.  Left panel shows the change to the accumulated azimuthal phase, $\delta\Phi_\phi$; right panel shows the change to the radial phase $\delta\Phi_r$.  Over a very large section of this parameter space, the change is quite large (dozens to thousands of radians).  Indeed, only within the narrow white band to the left of both panels do we have $|\delta\Phi| \le 1$.  It is worth noting that these dephasings scale proportional to $|\mathcal{R}|^2$, and inversely proportional to the system's mass ratio.}
\label{fig: dephasing pro} 
\end{figure*}

\begin{figure*}
\includegraphics[width=89mm]{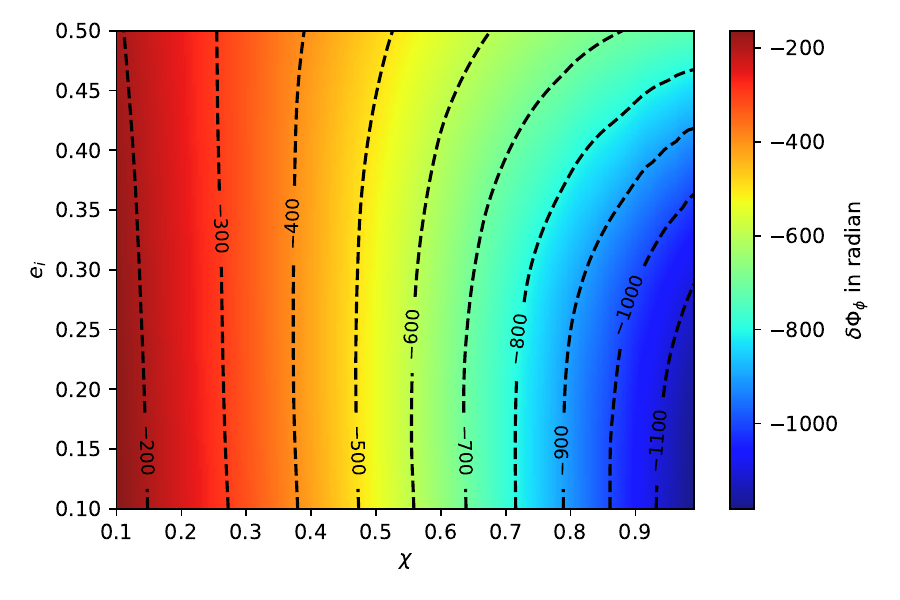}
\includegraphics[width=89mm]{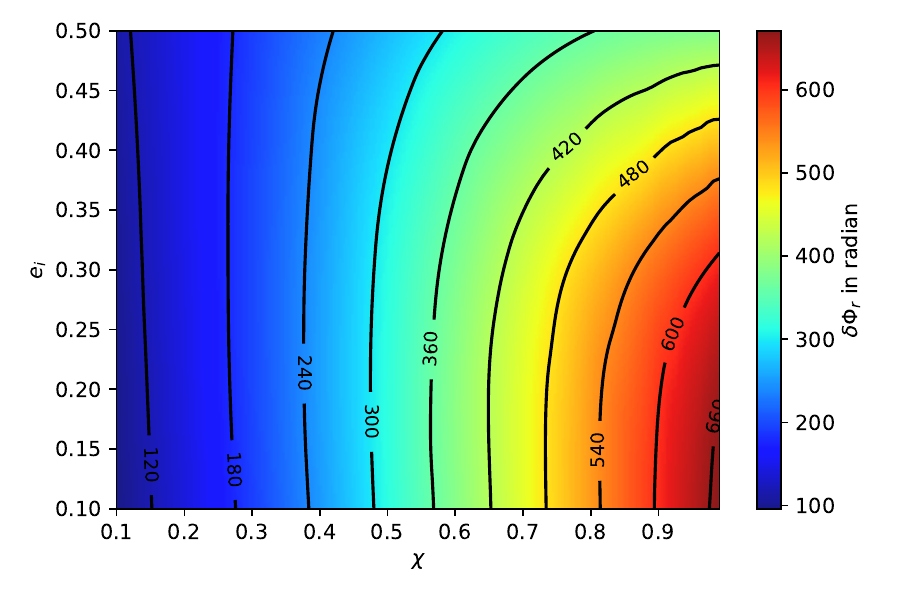}
\caption{Same as Fig.\ \ref{fig: dephasing pro} but for retrograde inspirals.  In contrast to prograde configurations, there is no place in this parameter space where the net dephasing is zero; we find non-negligible dephasing for all retrograde configurations.}
\label{fig: dephasing ret} 
\end{figure*}

At each point on the grid, we compute a large set of modes $Z^{\infty,H}_{lmn}$, including enough modes to ensure the fluxes (\ref{eq:dEIdt})--(\ref{eq:dLzHdt}) have reached numerical convergence: we stop computing modes when additional modes change these fluxes by less than $10^{-5}$.  See Ref.\ \cite{Hughes:2021exa} for a detailed discussion of our convergence criterion.  In this analysis, we use the fact that, during an adiabatic inspiral, the system evolves from geodesic to geodesic to deduce how an orbit's geometry evolves due to changes in an equatorial orbit's $E$ and $L_z$.  The ``geodesic evolves to geodesic'' constraint allows us to write down a Jacobian which relates the rates of change of $E$ and $L_z$ to the rates of change of $p$ and $e$:
\begin{eqnarray}
\left(\frac{dp}{dt}\right)^{\infty,H} &=& J_{pE}\left(\frac{dE}{dt}\right)^{\infty,H} + J_{pL_z}\left(\frac{dL_z}{dt}\right)^{\infty,H}\;,
\nonumber\\
\left(\frac{de}{dt}\right)^{\infty,H} &=& J_{eE}\left(\frac{dE}{dt}\right)^{\infty,H} + J_{eL_z}\left(\frac{dL_z}{dt}\right)^{\infty,H}\;.
\nonumber\\
\end{eqnarray}
The Jacobian entries $J_{pE}$--$J_{eL_z}$ can be read out of Appendix B of Ref.\ \cite{Hughes:2021exa}.

To construct an adiabatic inspiral, we construct a sequence of geodesics beginning at some initial point $[p(t_i), e(t_i)$], then integrating the fields $(dp/dt, de/dt)$ until we reach the edge of our data grid near the LSO.  Off grid data is found using two-dimensional cubic spline interpolation in the $p$ and $e$ directions.  Such interpolation is also used to assemble data for the waveform amplitude along the inspiral.  Additional data required for waveform construction, such as the geodesic frequencies $\Omega_r$ and $\Omega_\phi$ and their associated phases, are constructed at each geodesic orbit in this inspiral sequence.  We note that recent work \cite{Hughes2024} has shown that numerical errors can be reduced by integrating in $(dE/dt, dL_z/dt)$ and then using analytic mappings from $(E,L_z)$ to $(p,e)$ to characterize each orbit.  It may be interesting to revisit our analysis (which was largely completed before the analysis of Ref.\ \cite{Hughes2024} was done) taking this into account, but we are confident that the numerical errors introduced by integrating $(dp/dt, de/dt)$ do not have an important impact on our conclusions: the impact of tidal heating (our present focus) is independent of the systematic numerical errors examined in Ref.\ \cite{Hughes2024}.

\begin{figure*}
\includegraphics[width=89mm]{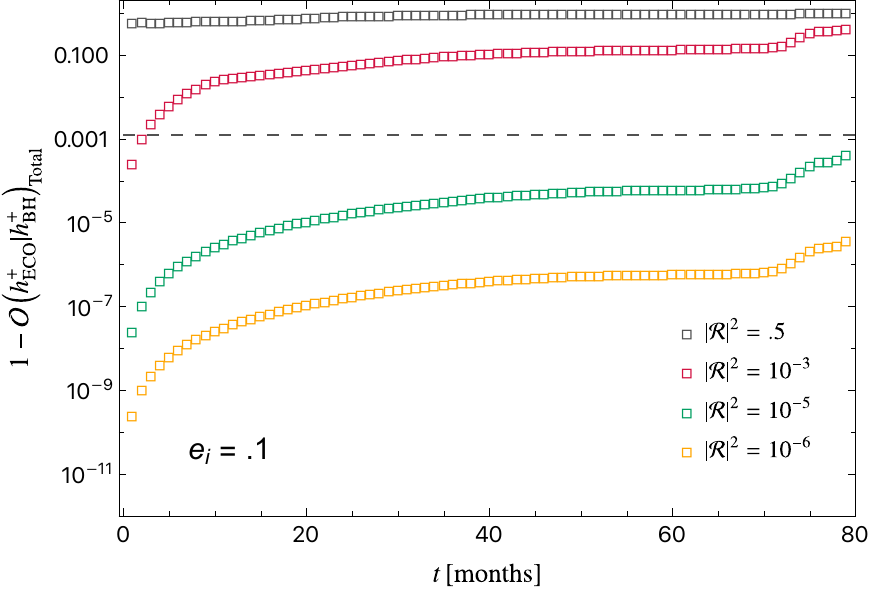}
\includegraphics[width=89mm]{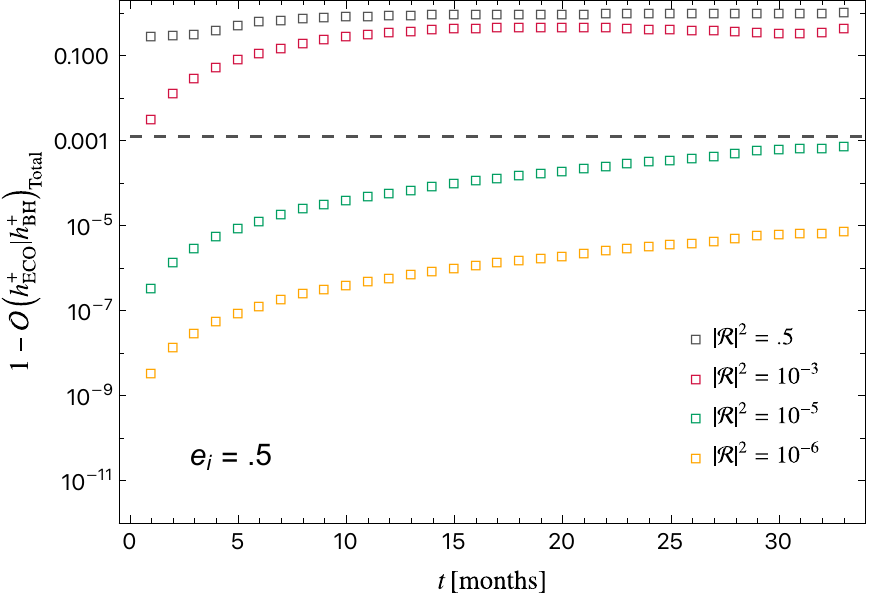}
\includegraphics[width=89mm]{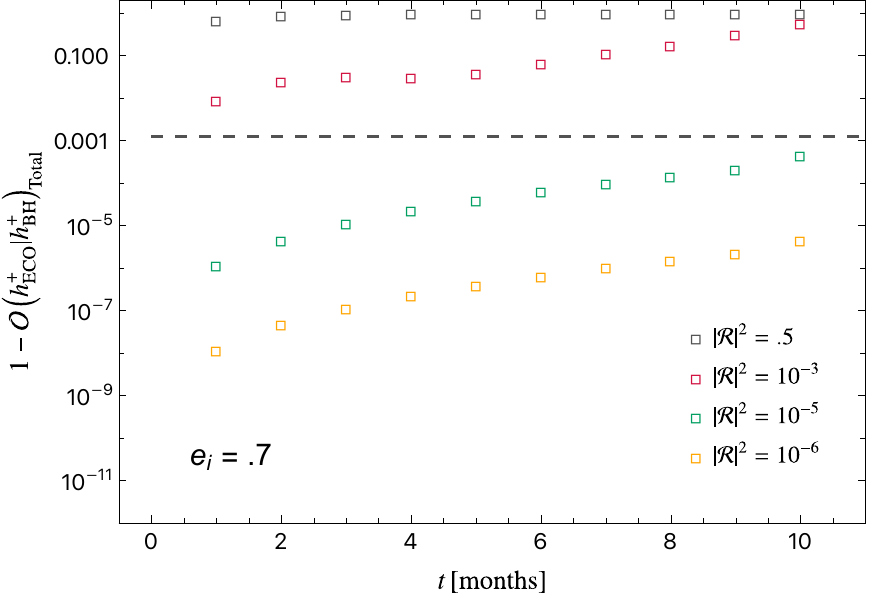}
\includegraphics[width=89mm]{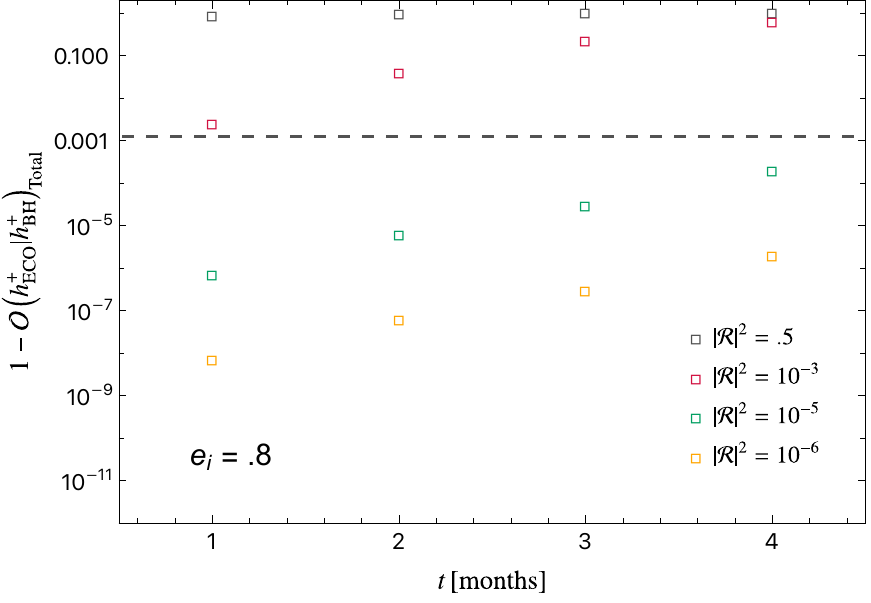}

\caption{Mismatch $\mathcal{M} = 1 - \mathcal{O}$ for prograde inspiral as a function of observation time between the plus polarization for a waveform computed with $|\mathcal{R}|^2 \neq 0$ (ECO) and another computed with $|\mathcal{R}|^2 = 0$ (BH), maximized over time and phase shift.  Each panel shows a different starting eccentricity; in all cases, the large BH has spin $\chi = 0.9$.  All other parameters are as described in the caption to Fig.\ \ref{fig:orbit change pro}.  The dashed horizontal line shows the threshold $\mathcal{M} = 1/(2\rho^2)$, with $\rho = 20$ chosen as a fiducial SNR associated with the true signal.  Notice that inspirals with reflectivity parameter $|\mathcal{R}|^2 = 10^{-3}$ exceed this threshold in all cases we consider here.}
\label{fig:mismatch} 
\end{figure*}

\section{Results}

We turn now to a discussion of our results, in particular how inspirals and their associated gravitational waveforms are changed when tidal heating is modified.  It is worth emphasizing that eccentric inspiral is changed by modified tidal heating in a way that is qualitatively different from what we find for quasi-circular inspiral: the trajectory that an inspiraling body follows in the $(p, e)$ plane changes as the reflectivity parameter $\mathcal{R}$ is changed.  In quasi-circular cases for fixed BH parameters, all inspirals enter our band at some initial radius $r_i$, and inspiral ends essentially at the innermost stable circular orbit.  The quasi-circular limit describes evolution through a 1-dimensional parameter space, and there is only one way to move through this space.  Eccentric inspiral, by contrast, involves an evolution through a sequence of orbits parameterized by $p$ and $e$; an adiabatic inspiral describes a curve $e(p)$ in this parameter space.  The slope of this curve, $de/dp$, depends only on the ratio $dE/dL_z = (dE/dt)/(dL_z/dt)$.  Modifying the tidal heating can modify this ratio, leading to an adiabatic inspiral following a different trajectory in the $[p(t),e(t)]$ than what we find in the pure-GR, $\mathcal{R} = 0$ case.

We begin by discussing how modifications to tidal heating alter the inspiral trajectory in the $(p, e)$ plane and then discuss how waveforms are changed.

\subsection{Modified trajectory}

Figure \ref{fig:orbit change pro} summarizes how the orbital trajectory changes for prograde orbits.  In each panel, we vary the spin of the massive BH (horizontal axis) and the initial eccentricity $e_i$ (vertical axis).  The starting value of semi-latus rectum $p$ is chosen so that the $m = 2$, $n = 0$ GW harmonic (which is typically the loudest voice contributing to the waveform) is at frequency $1$ millihertz.  Our results show the difference between inspirals computed with the ``normal'' tidal heating prediction of GR ($\mathcal{R} = 0$), and inspirals computed with no tidal heating at all ($\mathcal{R} = 1$).

The top panels of Fig.\ \ref{fig:orbit change pro} show how the duration of inspiral changes as we go from $\mathcal{R} = 0$ to $\mathcal{R} = 1$.  Top left shows the absolute change in duration $\Delta t$, and top right shows this as a percentage of the total inspiral time.  The total duration itself varies considerably over the parameter space, leading to rather different contour shapes between the two panels.

In both of the top panels of Fig.\ \ref{fig:orbit change pro}, there is a contour for which $\Delta t = 0$.  To the left of this contour, modifying tidal heating makes inspirals longer ($\Delta t > 0$); to the right, modifying tidal heating makes inspirals shorter ($\Delta t < 0$).  It is interesting that, for each starting eccentricity of prograde inspiral, there is some $\chi$, typically $\sim 0.1$ -- $0.3$, for which removing tidal heating does not change the duration of inspiral.  For many systems, $(dE/dt)^H$ changes sign at some point during inspiral.  The contour on which $\Delta t = 0$ shows where the impact of tidal heating before and after this change of sign balances out.  Over most of the parameter space, setting $\mathcal{R} = 1$ has a significant impact on the inspiral.  The effect is especially pronounced when $\chi \gtrsim 0.5$.  Inspiral goes deep into the strong field for large spins since the LSO is at smaller $p$; the impact of tidal heating is then quite strong.

The bottom two panels of Fig.\ \ref{fig:orbit change pro} show how changing $\mathcal{R}$ from $0$ to $1$ changes the values of $p$ and $e$ at which inspiral reaches the LSO.  (We show fractional changes in these parameters, $\Delta p_f/p_f$ and $\Delta e_f/e_f$; absolute changes in these parameters are shown in Appendix \ref{app: final params}.)  As with the inspiral duration, there is a contour along which $\Delta p_f$ and $\Delta e_f$ are zero.  For $e_i \lesssim 0.6$, this contour hovers near $\chi \approx 0.4$, though it goes to larger spin values for larger starting eccentricity.

The panels in Fig.\ \ref{fig: orbit change ret} show the same information as those in Fig.\ \ref{fig:orbit change pro}, but focusing now on retrograde orbits.  Note that we only consider $e_i \le 0.5$; retrograde orbits with $e_i > 0.5$ have very short durations with our starting requirement that $\omega_{20} = 2\pi \times 10^{-3}\,{\rm Hz}$.  One interesting difference compared to prograde orbits is that $(dE/dt)^H > 0$ across the parameter space of retrograde inspiral.  As such, changing from $\mathcal{R} = 0$ to $\mathcal{R} = 1$ always increases inspiral.  We also find that the final values $p_f$ and $e_f$ are always slightly larger with no tidal heating in retrograde cases.

\subsection{Example waveform and dephasing}

In Fig.\ \ref{fig: waveform}, we show examples of waveforms corresponding to inspiral using three different reflectivity values: $\mathcal{R} = 0$ (i.e., the standard GR value), $|\mathcal{R}|^2 = 0.5$, and $|\mathcal{R}|^2 = 10^{-3}$.  All three waveforms are for inspiral into a BH with $\chi = 0.9$, starting eccentricity $e_i = 0.8$, larger mass $M = 10^6\,M_\odot$, and mass ratio $M/\mu = 3\times10^{4}$.  The panels show roughly 45 minutes of $h_+$, starting 2 months after the signal enters band (left panel) and 3 months after entering band (right panel).  We can see a clear dephasing of the waveforms as $\mathcal{R}$ is increased from zero.  At two months, the inspiral with $|\mathcal{R}|^2 = 10^{-3}$ is essentially in phase with the GR waveform, though the one with $|\mathcal{R}|^2 = 0.5$ is highly dephased from the other two.  At three months, a clear separation can be discerned even between $\mathcal{R} = 0$ and $|\mathcal{R}|^2 = 10^{-3}$.

A crude but commonly applied rule of thumb is that a dephasing $\delta\Phi \sim 1$ radian between two model waveforms indicates that the models can be clearly separated \cite{Lindblom:2008cm}.  In our case, this would imply that the value of $|\mathcal{R}|^2$ which yields dephasing $\delta\Phi \sim 1\,{\rm rad}$ is the minimum horizon reflectivity which may be discerned from the GR prediction $\mathcal{R} = 0$.  It should be strongly emphasized that such a rule of thumb should be supplemented by a more careful parameter study.  Correlations between parameters might hide the effect of non-zero $\mathcal{R}$; non-zero $\mathcal{R}$ could perhaps be found using a ``normal'' GR waveform model at the cost of some systematic parameter error.  Only a thorough study can determine the true threshold one should apply for $\mathcal{R}$.

With this caution in mind, Figs.\ \ref{fig: dephasing pro} and \ref{fig: dephasing ret} illustrate the maximum dephasing we find as functions of BH spin $\chi$ (horizontal axis) and initial eccentricity $e_i$ (vertical axis).  Figure \ref{fig: dephasing pro} illustrates prograde inspiral; Fig.\ \ref{fig: dephasing ret} is for retrograde.  In both figures we compare the GR case $\mathcal{R} = 0$ with maximum reflectivity $\mathcal{R} = 1$.  Left panel in both figures shows the change to the axial phase $\Phi_\phi$ found by integrating the axial frequency $\Omega_\phi$ over inspiral; right panel shows the change to the radial phase $\Phi_r$ found by integrating $\Omega_r$.  All the various voices which contribute to the waveform involve harmonics of these phases: $\Phi_{mn} = m\Phi_\phi + n\Phi_r$.

We find the largest dephasing for high spin, low eccentricity, prograde inspirals; these are also the cases for which inspiral has the largest number of cycles in band.  As we found with various parameters, for prograde inspirals there exists a contour at which the dephasing vanishes; in Fig.\ \ref{fig: dephasing pro}, we highlight where $-1 \le \delta\Phi_x \le 1$ (for $x = \phi$ and $x = r$) as thin white strips near the left-hand side of the two panels.  No such contour exists for retrograde inspirals.  All regions outside this strip have dephasing with magnitude larger than this; indeed, for both prograde and retrograde inspirals, the maximum dephasing reaches hundreds to thousands of radians across much of the parameter space, though tending to be larger (by about an order of magnitude) for prograde inspirals.  Finally, it is worth bearing in mind that the dephasing scales with $|\mathcal{R}|^2$, and inversely with the system's mass ratio.  Although a more careful and comprehensive analysis is needed to truly ascertain how sensitive EMRI waveforms are to changes in tidal heating, the very large dephasings we find here show how sensitive these waveforms are to this effect, and indicate the promise of these measurements as a test of BH event horizons.

\subsection{Mismatch}

As a supplement to our dephasing analysis, we also examine the {\it mismatch} between waveforms with $\mathcal{R} = 0$ and $\mathcal{R} \ne 0$, as was done in Ref.\ \cite{Datta:2019epe}.  We begin by defining the overlap between two waveforms, which provides a robust indicator of how well two waveforms agree in the presence of an expected noise background.  The overlap is given by

\begin{equation}\label{eq: overlap}
\mathcal{O}(h_1|h_2) = \frac{\left\langle h_1|h_2\right\rangle}{\sqrt{\left\langle h_1|h_1\right\rangle \left\langle h_2|h_2\right\rangle}}\,,
\end{equation}
where the noise-weighted inner product $\langle h_1|h_2 \rangle$ is
\begin{equation}
\left\langle h_1|h_2\right\rangle = 4\Re\,\int_{0}^{\infty} \frac{\tilde{h}_1 \tilde{h}^*_2}{S_n(f)} df\,.
\end{equation}
Here $S_n(f)$ is the one-sided noise power spectral density (PSD), $\tilde{h}_1(f)$ and $\tilde{h}_2(f)$ represent the Fourier transforms of the waveforms $h_1(t)$ and $h_2(t)$, and the superscript $^*$ denotes complex conjugate.  Since the waveforms are defined up to an arbitrary time and phase shift, we maximize the overlap over these parameters, as described in Ref.~\cite{Datta:2019epe}.  For the PSD, we use the LISA curve including the confusion background from unresolved Galactic stellar binaries over a one-year mission lifetime, following from Ref.\ \cite{Cornish:2018dyw}.  

The overlap is defined such that $\mathcal{O} = 1$ indicates perfect agreement between two waveforms.  The mismatch $\mathcal{M}$ is defined as
\begin{equation}
    \mathcal{M} = 1 - \mathcal{O}\;,
\end{equation}
and indicates the degree to which two waveforms are not in agreement.  In Fig.~\ref{fig:mismatch} we show the mismatch for the plus polarization of waveforms of prograde EMRIs with different reflectivities.  We fix $\chi = 0.9$ for all cases; each panel shows a different value of initial eccentricity $e_i$.  The mismatch is shown as a function of observation time for orbits starting at $1{\rm mHz}$.  Since we start each inspiral when $\omega_{20} = 2\pi \times 10^{-3}\,{\rm sec}^{-1}$, the inspiral duration varies significantly depending on the initial eccentricity.

For small values of $|\mathcal{R}|^2$, the mismatch during the initial months is very small but it grows rapidly with time. Interestingly, the accumulated mismatch when the systems approach the LSO is of the same order-of-magnitude for different initial eccentricities, even though the inspiral time is much shorter for larger initial eccentricities. This is related to the fact that the impact of tidal heating tends to be more important for larger eccentricities, compensating for the smaller number of cycles.  This feature can also be seen in the dephasing plot, Fig.\ \ref{fig: dephasing pro}. As expected, the mismatch decreases with decreasing $|\mathcal{R}|^2$.  As discussed in the next section, these results suggest that reflectivities as small as $|\mathcal{R}|^2 \sim 10^{-5} $ could have an observable impact, in agreement with previous results obtained for circular orbits~\cite{Datta:2019epe}, though we emphasize a proper parameter study must be done to assess this suggestion.  We find similar results for retrograde orbits, as shown in Appendix~\ref{app: retro mismatch}.

\section{Discussion}

It has long been established that EMRIs will be unparalleled probes of fundamental physics and unique sources for the LISA mission~\cite{Barausse:2020rsu, LISA:2022kgy,LISAConsortiumWaveformWorkingGroup:2023arg,Colpi:2024xhw}; see also \cite{Maselli:2021men, Cardoso:2022whc, Rahman:2022fay, Rahman:2023sof, AbhishekChowdhuri:2023gvu} for specific examples of recent, detailed studies of this type.  EMRI dynamics are affected by modified boundary conditions at a compact object's surface, which give rise to modified tidal heating, modified fluxes, and resonant QNM excitations in a consistent fashion.

In the spirit of devising a model independent test of the presence of a Kerr horizon, we have studied the signal emitted by a small body in eccentric orbits around a Kerr BH.  In order to formulate such a test, we have focused our analysis on a simplified modeling of the tidal heating, leaving other effects like modified boundary conditions and possible resonances for future work.  Our approach can be regarded as a simple modification of the standard BH case, modifying adiabatic inspiral by adjusting the down-horizon flux which describes tidal coupling of an orbiting body to the primary's event horizon.

Two waveforms can be considered indistinguishable if the mismatch $\mathcal{M}$ between them satisfies $\mathcal{M} \lesssim 1/2\rho^2$ \cite{Flanagan:1997kp, Lindblom:2008cm}, where $\rho$ is the SNR.  For an EMRI with $\rho \approx 20$, this implies a ``threshold'' mismatch $\mathcal{M} \lesssim 10^{-3}$ for two signals to be indistinguishable.  In Figs.\ \ref{fig:mismatch} and \ref{fig:mismatch retro}, we show that at $\rho = 20$, this level of mismatch is exceeded across a very wide range of plausible EMRI parameters.

Previous work \cite{Datta:2019epe} has shown that this level of mismatch is exceeded even for small BH spins.  The present work shows this phenomenon also for eccentric inspirals, suggesting that horizon reflectivity $|\mathcal{R}|^2$ can be constrained to quite small values by future measurements of these events.  We note that as eccentricity increases, inspiral time tends to decrease, resulting in less time for the effect of tidal heating to accumulate.  We nevertheless find that the impact of $|\mathcal{R}|^2$ on an inspiral is similar to that for circular orbits, highlighting the fact that the effect of ``down-horizon'' flux on the inspiral is enhanced for nonzero eccentricities.  Such an effect has been noted in analytical studies as well \cite{Datta:2023wsn, Munna:2023vds}.

As a result, for a supermassive object with $\chi = 0.9$ and a signal having $\rho = 20$, we can infer a very stringent constraint on the reflectivity $|\mathcal{R}|^2 \sim \mathcal{O}(10^{-5})$ based on the results shown in Figs.~\ref{fig:mismatch} and~\ref{fig:mismatch retro}. This implies that an EMRI detection is capable of discerning an effective reflectivity of the central supermassive object as small as $\sim \mathcal{O}(0.001)\%$. This outcome aligns with previous findings in the context of circular orbits. It underscores the importance of accurately modeling tidal heating for all types of orbital configurations to prevent substantial dephasing and systematic errors. Furthermore, we have demonstrated that the inclusion or absence of tidal heating can serve as a robust, model-independent discriminator for the presence of a horizon in the central supermassive object, even in eccentric orbits.

Our analysis exclusively focuses on the modification of fluxes at the leading order in the mass ratio, specifically considering only the leading-order dissipative component of the self-force~\cite{Barack:2009ux, Cardoso:2011xi}. We have neglected conservative contributions and higher-order terms. Although these corrections are crucial for parameter estimation, their effects are unlikely to be confused with those of tidal heating. Tidal heating effects are typically much more pronounced, especially for realistic spin values when $\mathcal{R}$ is not negligibly small. Consequently, our earlier findings under circular orbit assumptions are substantiated, suggesting that accurate constraints can be attained by modeling the (partial) absence of tidal heating in state-of-the-art waveform approximations at the leading order.

In Refs.~\cite{Sago:2021iku, Maggio:2021uge, Cardoso:2022fbq}, alongside the consideration of tidal heating, the role of different boundary conditions and resonances resulting from the excitation of low-frequency quasinormal modes, which are prevalent for ECOs~\cite{Pani:2010em, Macedo:2013jja, Cardoso:2019rvt}, was also explored. This investigation demonstrated that modified boundary conditions, and in particular resonances, significantly amplify the effects, enhancing the sensitivity to even smaller reflectivities. In the current work, we have observed that the majority of the conclusions obtained for circular orbits extend to eccentric cases. Therefore, we anticipate that the conclusions drawn in the circular orbit analysis of Refs.~\cite{Sago:2021iku, Maggio:2021uge} are also likely to apply to eccentric orbits. However, it is important to note that Ref.~\cite{Cardoso:2022fbq} highlighted that a frequency domain analysis tends to overestimate the impact of resonances, emphasizing the need for further investigations in this regard.

In addition to our model-agnostic approach, an accurate modelling of the effect of tidal heating relies on the properties of the ECOs. In this context, the presence of structure at the horizon scale (even possibly arising from quantum corrections) can also have an impact on tidal heating. If the area or mass of a BH is quantized then it can also significantly modify the features of tidal heating and its signatures~\cite{Agullo:2020hxe, Datta:2021row, Chakravarti:2021jbv}.
Additionally, in Ref.~\cite{Datta:2020rvo}, apart from examining the reflectivity, the impact of the reflective surface position $(r_s)$, denoted as $r_s = r_+(1+\varepsilon)$ where $\varepsilon<1$, was analyzed analytically and later extended in Ref.~\cite{Chakraborty:2021gdf}. This analysis revealed a degeneracy between the reflectivity and $\varepsilon$, establishing that $\varepsilon$ on the order of $\sim \mathcal{O}(|\mathcal{R}|^2)$ can have a noticeable impact. This suggests an observable effect of $\varepsilon$ on the order of $\mathcal{O}(10^{-5})$. 
Using a frequency-domain approach, Ref.~\cite{Maggio:2021uge} further demonstrated that the impact of resonances and ab-initio boundary conditions can contribute to make even smaller corrections detectable. If the current work serves as an indicator, it implies the possibility of retaining these findings in eccentric cases as well. Therefore, the impact of $\varepsilon$ should be investigated meticulously in the context of EMRIs, possibly in a model-dependent fashion.

\begin{figure*}
\includegraphics[width=89mm]{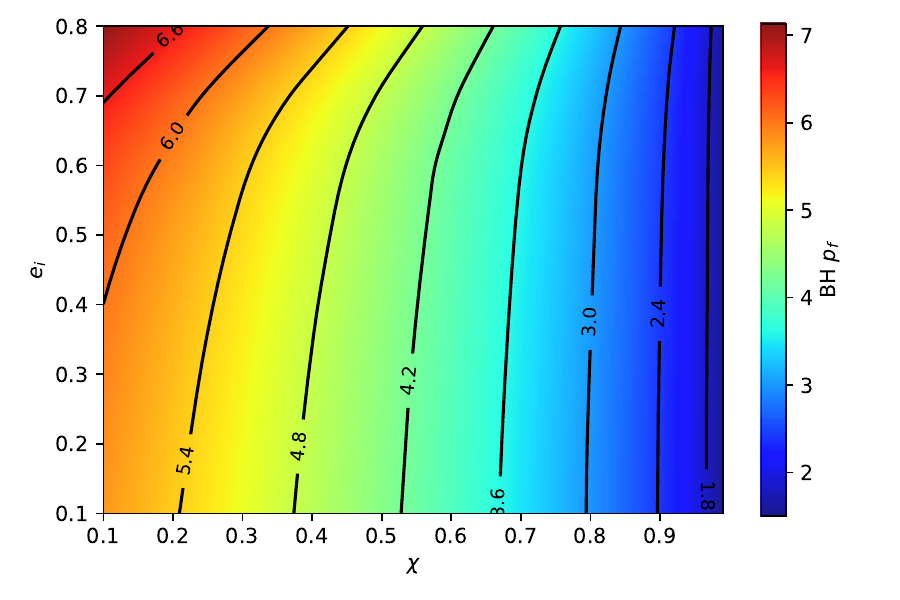}
\includegraphics[width=89mm]{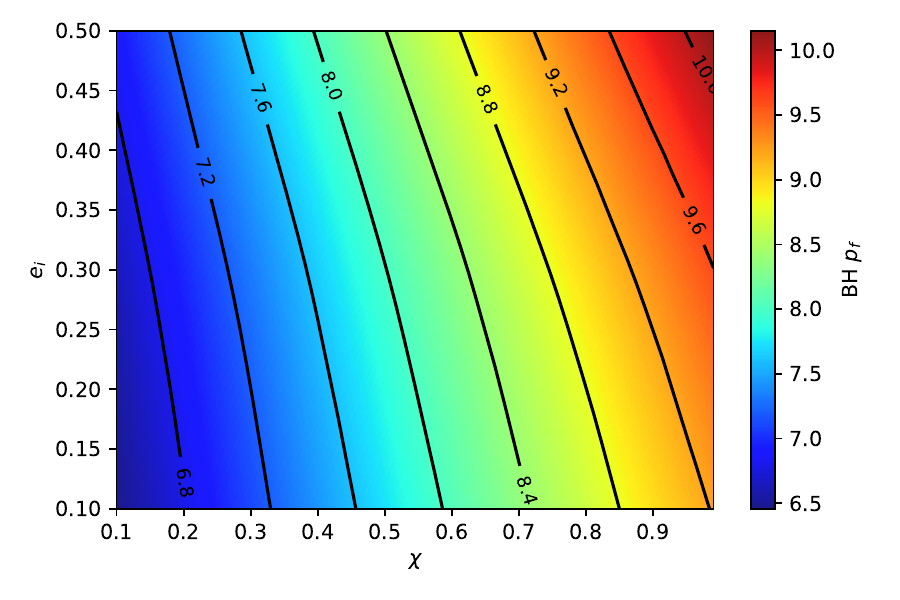}
\includegraphics[width=89mm]{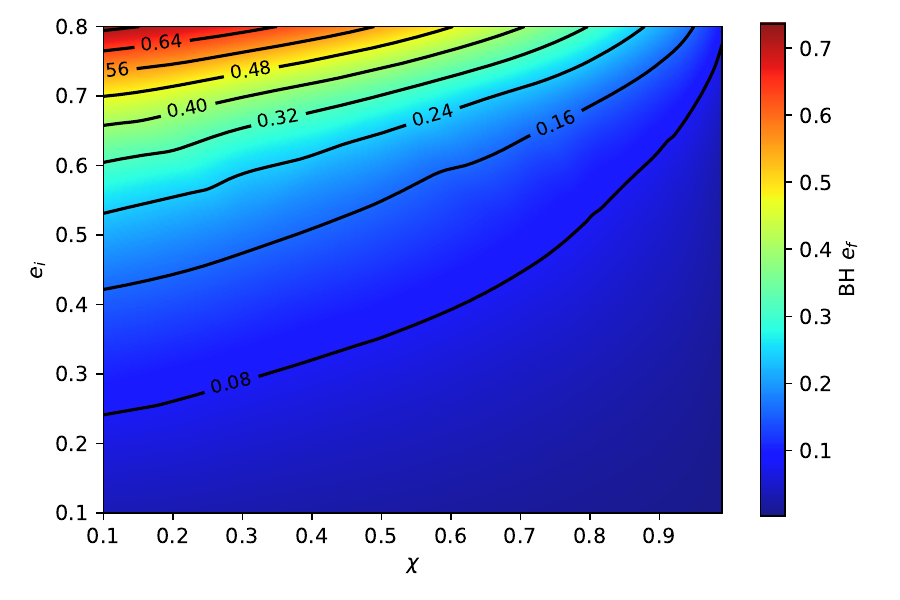}
\includegraphics[width=89mm]{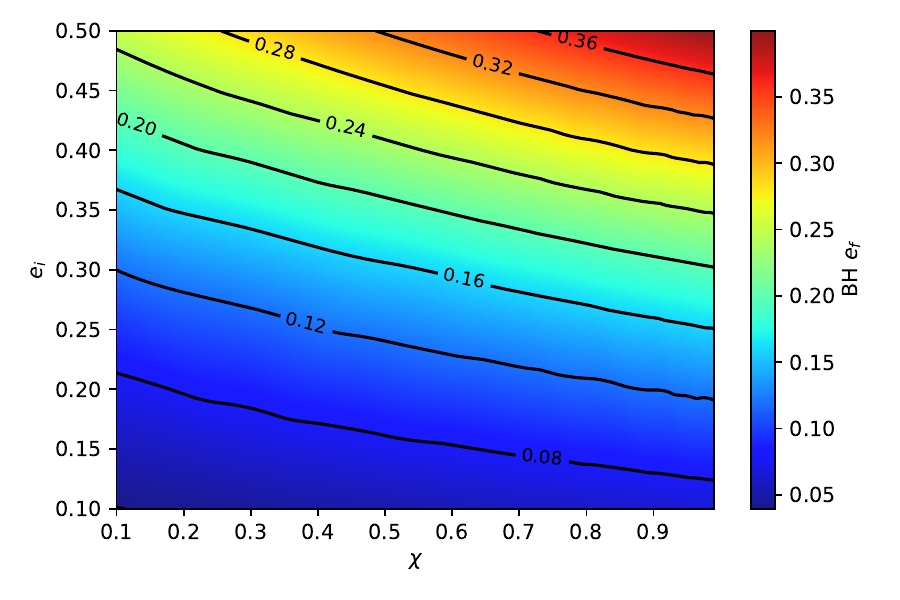}
\includegraphics[width=89mm]{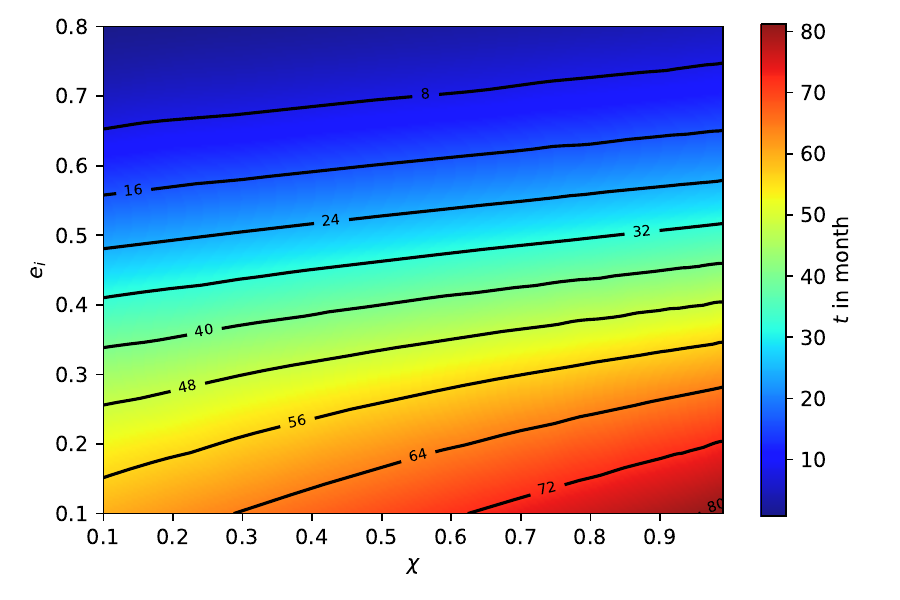}
\includegraphics[width=89mm]{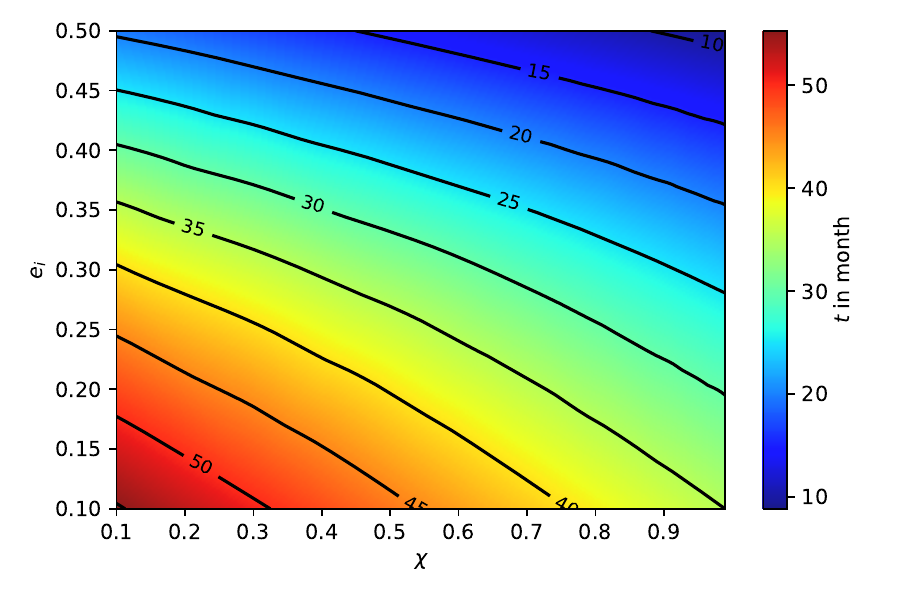}
\caption{The characteristics of inspiral when the central object is assumed to be a BH, i.e., when we put $\mathcal{R} = 0$.  Each panel shows some important characteristic of the inspiral for a system that begins with spin $\chi$ (horizontal axis) and starting eccentricity $e_i$ (vertical axis), and for which $M = 10^6\,M_\odot$, $M/\mu = 3\times10^4$.  All inspirals begin when the GW frequency corresponding to the $m = 2$, $n = 0$ voice crosses $10^{-3}$ Hz.  Left-hand panels show results for prograde inspirals; right-hand panels are for retrograde.  From top to bottom, the panels show the final value of semi-latus rectum, $p_f$, when inspiral ends; the final value of eccentricity $e_f$; and the total inspiral duration in months.  Note that our retrograde data covers a smaller span of initial eccentricity than the prograde data, since high eccentricity retrograde inspirals have very short duration with the above constraints.}
\label{fig: BH final parameters} 
\end{figure*}

\begin{figure*}
\includegraphics[width=89mm]{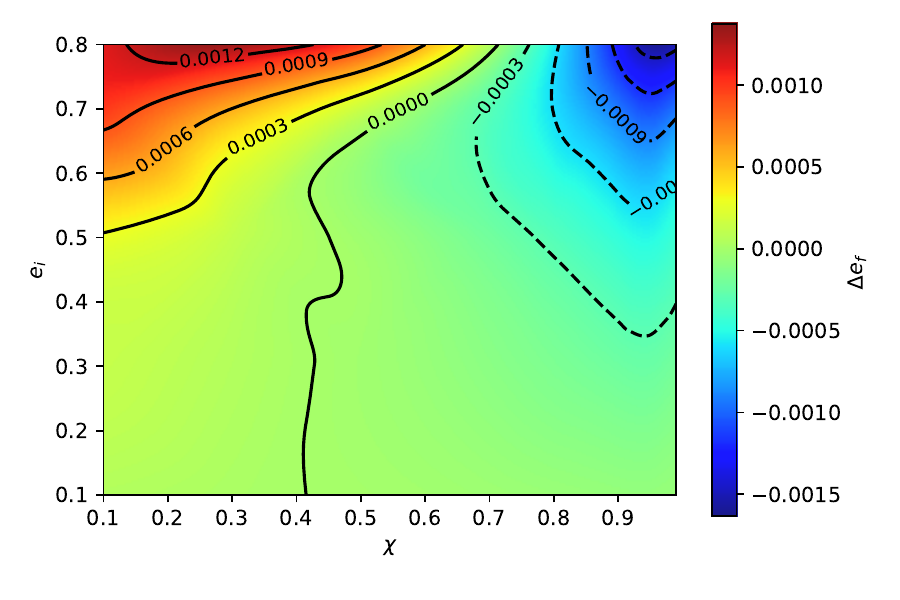}
\includegraphics[width=89mm]{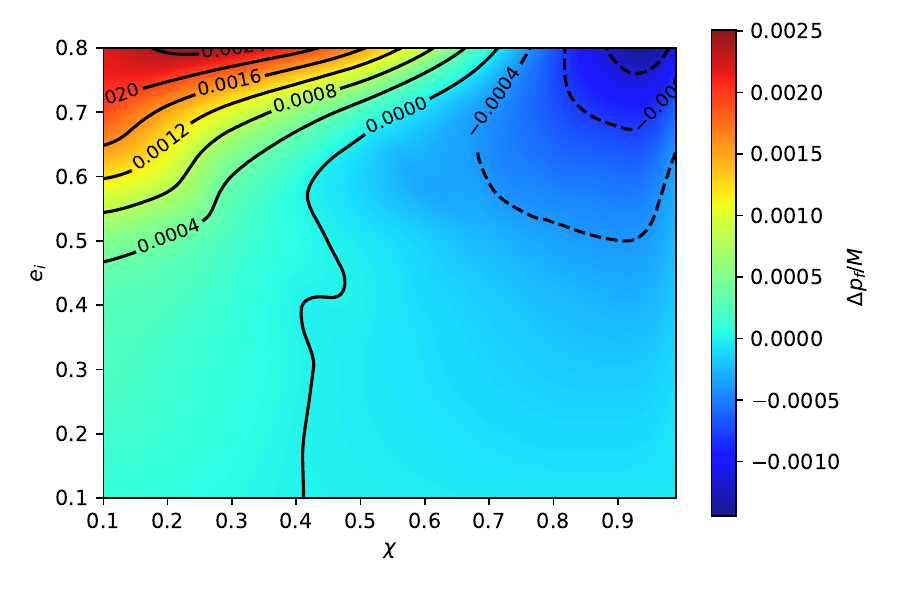}
\includegraphics[width=89mm]{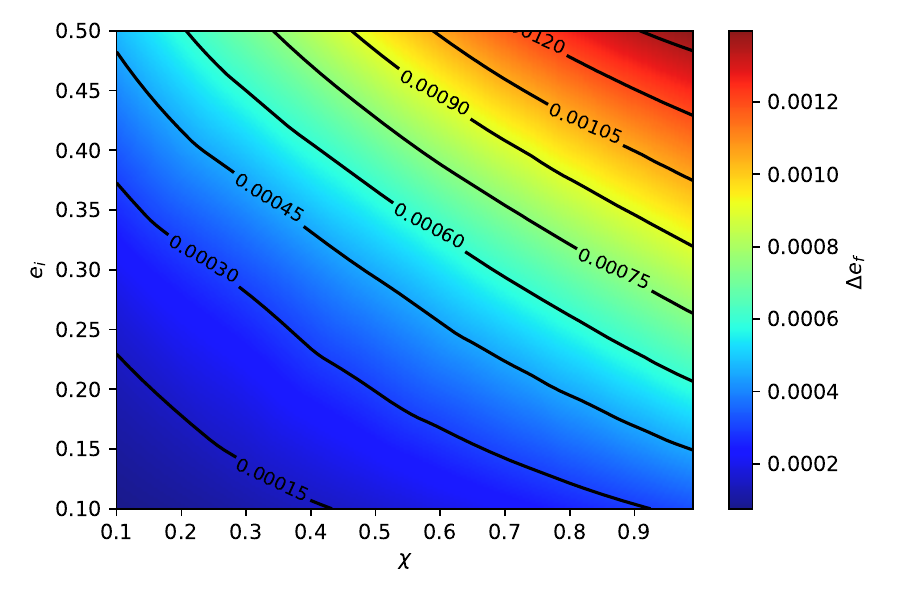}
\includegraphics[width=89mm]{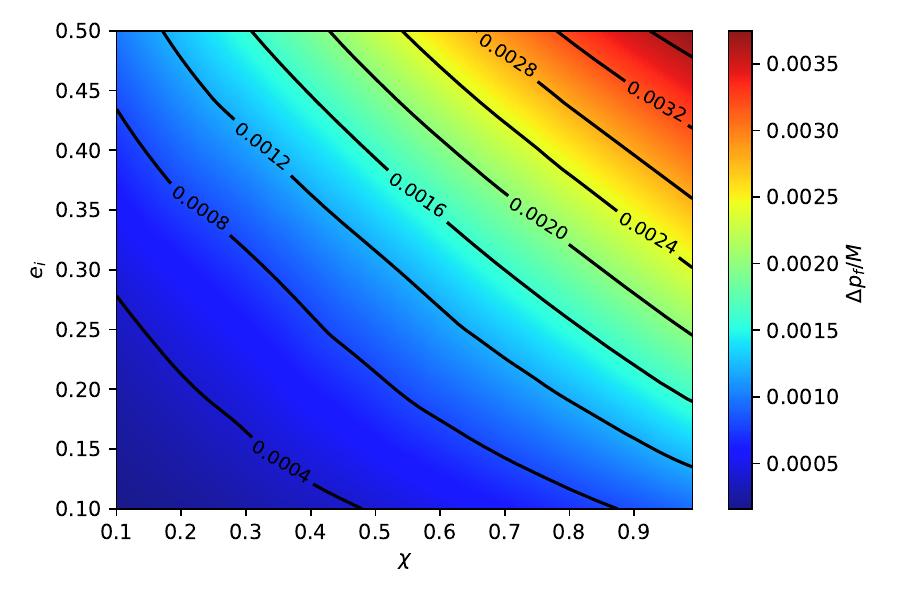}
\caption{The absolute change in final eccentricity $e_f$ (left-hand panels) and final semi-latus rectum $p_f$ (right) comparing a pure GR inspiral ($\mathcal{R} = 0$) with maximum reflectivity ($\mathcal{R} = 1$).  Top is for prograde inspiral; bottom is for retrograde.  These plots complement the fractional changes shown in Figs.\ \ref{fig:orbit change pro} and \ref{fig: orbit change ret}.}
\label{fig: e p change} 
\end{figure*}

Another crucial point is the connection between tidal deformability and tidal heating. In Refs.~\cite{Pani:2019cyc,Datta:2021hvm, Piovano:2022ojl} it has already been demonstrated the observability of very small values of tidal deformabilities with circular equatorial EMRI which was extended to eccentric orbits in Ref.~\cite{Bernaldez:2023xoh, Zi:2023pvl}. The connection between reflectivity and tidal deformability has already been established in several works~\cite{Nair:2022xfm, Chakraborty:2023zed}. Therefore, a holistic approach considering both tidal effects can not only break the degeneracy between $|\mathcal{R}|^2$ and $\varepsilon$ but also may lead to better measurement of reflectivity altogether. These aspects must be explored in the future.

It is worth emphasizing that our analysis is grounded in the assumption that the external geometry of the central object can be adequately described by the Kerr metric. ECOs may potentially exhibit deviations from the Kerr metric~\cite{Raposo:2018xkf, Barcelo:2019aif, Bena:2020see,Bianchi:2020bxa, Bena:2020uup, Bianchi:2020miz,Bah:2021jno}. Such departures from the Kerr metric influence the dissipative aspects of EMRIs at the leading order in the mass ratio and as a consequence provide an additional avenue for distinguishing the presence of horizons.

%%%%%%%%%%%%%%%%%%%%%%%%%%%%%%%%%%
\begin{acknowledgments}
%%%%%%%%%%%%%%%%%%%%%%%%%%%%%%%%%%
R.B. acknowledges financial support provided by FCT - Funda\c c\~{a}o para a Ci\^{e}ncia e a Tecnologia, I.P., under the Scientific Employment Stimulus -- Individual Call -- Grant No. 2020.00470.CEECIND and under Project No. 2022.01324.PTDC.
SD gratefully acknowledges the use of the IUCAA computing cluster, Sarathi, and Albert-Einstein Institute (AEI) cluster Atlas.
SAH was supported by NASA ATP Grant 80NSSC18K1091 and NSF Grant PHY-2110384; TK was supported by ATP Grant 80NSSC18K1091
while at MIT.
PP is partially supported by the MUR PRIN Grant 2020KR4KN2 ``String
Theory as a bridge between Gauge Theories and Quantum Gravity'' and by the MUR FARE programme (GW-NEXT, CUP:~B84I20000100001). 
S.D. would like to thank Sudhagar S. for useful discussions related to handling computational runs.
%%%%%%%%%%%%%%%%%%%%%%%%%%%%%%%%%%
\end{acknowledgments}
%%%%%%%%%%%%%%%%%%%%%%%%%%%%%%%%%%

\appendix

\section{BH results}
\label{app:BH results}

For completeness, in this Appendix we show the results for an inspiral process when the central object is a BH ($\mathcal{R}=0$). This is shown in Fig.~\ref{fig: BH final parameters}, where we present the semilatus rectum (upper panels), the eccentricity (central panels) and inspiral time (bottom panels), all computed at the LSO. The results for prograde motion are displayed in the left column, while the outcomes for retrograde motion are shown in the right column. All the binaries were initiated at a semi-latus rectum $p$ such that $\omega_{20} = 2\pi \times 10^{-3}\,{\rm sec}^{-1}$. The initial eccentricity of the system and the spin of the central object are plotted in the $x$ and $y$ axes, whereas the colored contours represent the value of the relevant parameter for such configurations.

\section{Absolute differences in the final parameters}
\label{app: final params}

In the main text, we discussed the changes in several orbital quantities if tidal heating is neglected. The results demonstrated that the inspiral takes a different trajectory if tidal heating is modified. For relative comparison with the standard BH case, we only showed the fractional differences between the cases where tidal heating is neglected or not. To complement that information here we also show the absolute differences $\Delta e_f$ and $\Delta p_f$, see Fig.~\ref{fig: e p change}. The top panels show results for prograde orbits, whereas in the bottom panels, we show results for retrograde orbits. As discussed in the main text, for prograde orbits there is a contour line where the difference between the BH case ($|\mathcal{R}|=0$) and the ECO case ($|\mathcal{R}|=1$) vanishes.

\section{Mismatch for retrograde orbits}
\label{app: retro mismatch}

\begin{figure*}
\includegraphics[width=89mm]{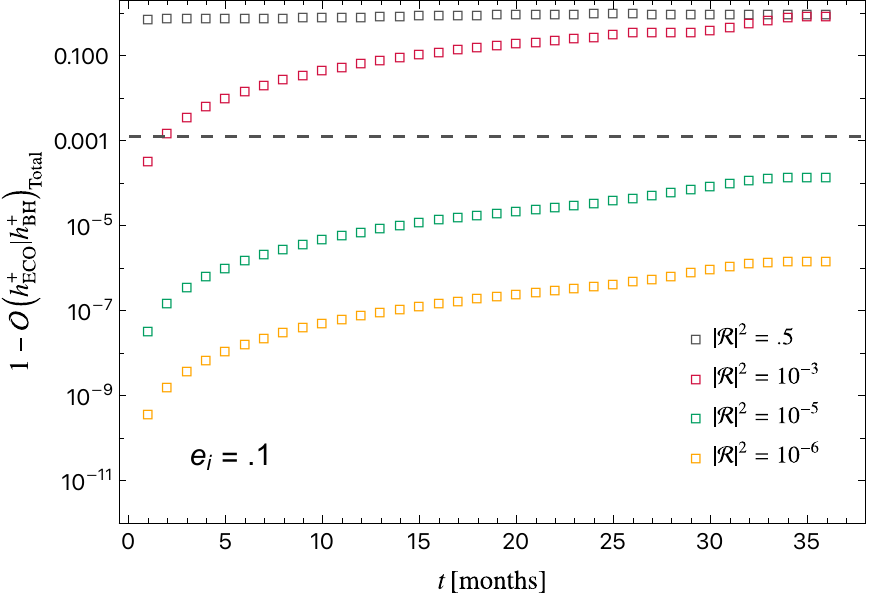}
\includegraphics[width=89mm]{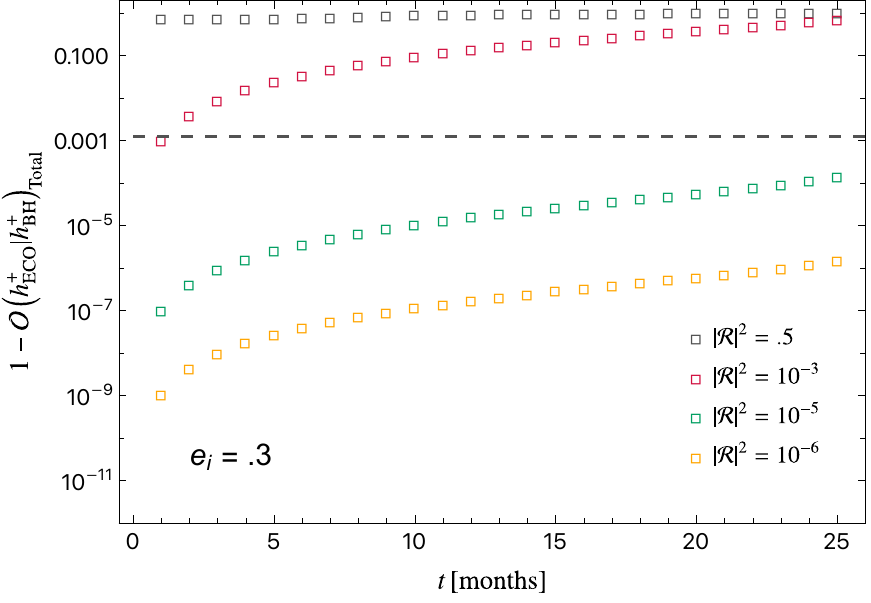}
\includegraphics[width=89mm]{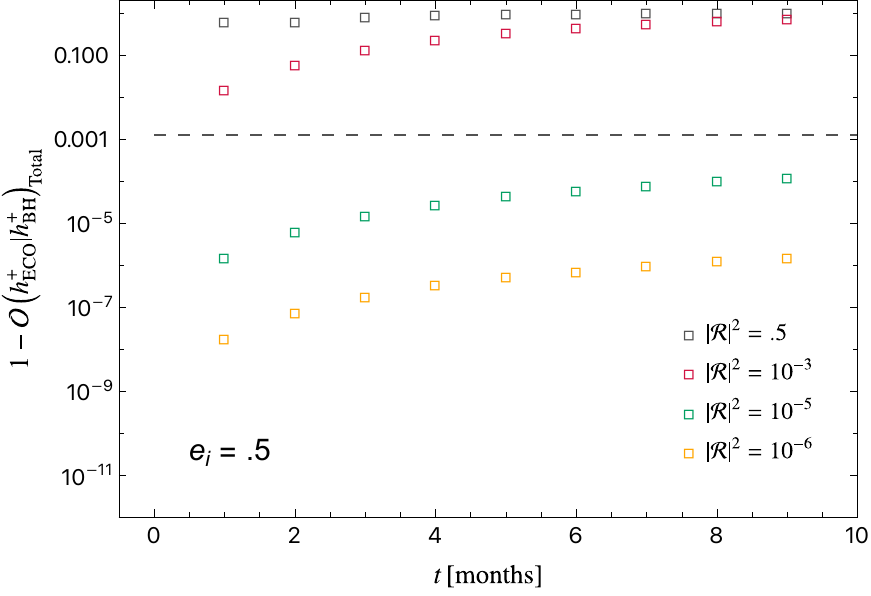}
\caption{Mismatch $\mathcal{M} = 1 - \mathcal{O}$ for retrograde inspiral; otherwise the same as Fig.\ \ref{fig:mismatch}. Notice that, just as in the prograde examples, inspirals with reflectivity parameter $|\mathcal{R}|^2 = 10^{-3}$ exceed the $10^{-3}$ mismatch threshold in all cases we consider here.}
\label{fig:mismatch retro} 
\end{figure*}

In this section, we discuss the mismatch results for retrograde orbits. In Fig.~\ref{fig:mismatch retro} we show the mismatch $\mathcal{M} \equiv 1 - \mathcal{O}$ for the waveforms of retrograde EMRIs with different reflectivities while fixing $\chi=0.9$ and the initial eccentricity $e_i$, each plot corresponding to a different $e_i$ similarly to the plots shown in Fig.~\ref{fig:mismatch} for prograde orbits. The mismatch is shown as a function of observation time for orbits starting at $1\,{\rm mHz}$. 
Similar to the prograde orbits, for smaller values of $|\mathcal{R}|^2$, the mismatch during the initial months is small and it grows with time. As expected, the mismatch decreases with decreasing $|\mathcal{R}|^2$. From these results, one can deduce that, similarly to the prograde orbits case, reflectivities as small as $|\mathcal{R}|^2 \sim 10^{-5} $ could have an observable impact.

\bibliography{References}

\end{document}